\documentclass[11pt]{article}
\usepackage{axodraw}
\usepackage{amsfonts}
\usepackage{epsfig}
\usepackage{amssymb}
\newcommand{\mrm}[1]{\mbox{\rm #1}}
\newcommand{\nn}{\nonumber}
\newcommand{\be}{\begin{equation}}
\newcommand{\bea}{\begin{eqnarray}}
\newcommand{\eea}{\end{eqnarray}}
\newcommand{\ee}{\end{equation}}

\newcommand{\eq}[1]{Eq.~(\ref{#1})}
\newcommand{\D}{D\hspace{-8pt}\slash}
\newcommand{\dv}{\partial\hspace{-7pt}\slash}
\newcommand{\Dl}{\stackrel{\hspace{3pt}\leftarrow}{D\hspace{-8pt}\slash}}

\newcommand{\Dr}{\stackrel{\hspace{3pt}\rightarrow}{D\hspace{-8pt}\slash}}
\newcommand{\dvr}{\stackrel{\hspace{3pt}\rightarrow}{\partial\hspace{-7pt}\slash}}

\begin{document}
\thispagestyle{empty}
\begin{flushright}
{\tt hep-ph/0406019}\\
{FTUAM-04-11}\\
{IFT-UAM/CSIC-04-27} \\
{UCSD/PTH 04-07}

\end{flushright}
\vspace*{1cm}
\begin{center}
{\Large{\bf  Renormalization of Lepton Mixing for Majorana Neutrinos} }\\
\vspace{.5cm}
A. Broncano$^{\rm a,}$\footnote{alicia.broncano@uam.es}, 
M.B. Gavela$^{\rm a,}$\footnote{gavela@delta.ft.uam.es} and 
Elizabeth Jenkins$^{\rm b,}$\footnote{ejenkins@ucsd.edu}
 
\vspace*{1cm}
$^{\rm a}$ Dept. de F\'{\i}sica Te\'orica, C-XI, and 
IFT, C-XVI, Facultad de Ciencias, \\ 
Univ. Aut\'onoma de Madrid, Cantoblanco, 28049 Madrid, Spain \\
$^{\rm b}$ Dept. of Physics, University of California at San Diego, 
9500 Gilman Drive, La Jolla, CA 92093, USA

\vspace{.3cm}

%
%
\begin{abstract}

 We discuss the one-loop electroweak renormalization of the leptonic mixing 
 matrix in the case of Majorana neutrinos, and establish
its relationship with the renormalization group evolution
 of the dimension five operator responsible for the light Majorana
 neutrino masses. We compare our results in the effective theory with those
in the full seesaw theory.

\end{abstract}

\end{center}
%
%

\pagestyle{plain} 
\setcounter{page}{1}
\setcounter{footnote}{0}
\setcounter{section}{0}

\section{Introduction}

The complete renormalization of the Standard Model (SM) includes the 
renormalization of the mixing angles and phases of the CKM matrix. 
Given the experimental evidence for neutrino oscillations,  
leptonic mixing occurs, and its pattern becomes part of the flavor puzzle.
In the lepton sector,  
the renormalization group evolution of the flavor-mixing parameters may be
quantitatively more significant than in the quark 
sector, as leptons mix with large intergenerational angles and  
the dominant gauge interactions are the electroweak interactions, which involve flavor mixing.
In contrast, in the quark sector, the CKM matrix is close to diagonal and the flavor-blind
strong interactions dominate. 
In this work, we discuss the relationship between the one-loop electroweak 
renormalization of the leptonic mixing parameters and the 
renormalization group equations (RGE) for higher-dimension leptonic operators of the unbroken and spontaneously broken effective theories.

Neutrino masses can be Dirac or 
Majorana.  The Majorana case is natural in the technical sense of `t~Hooft~\cite{natural}, and is
favored theoretically.  
Neutrinos are massless in the SM, since there are no right-handed neutrinos and the theory conserves $(B-L)$, where $B$ and $L$ denote baryon and lepton number, respectively.
Non-null neutrino masses are a signal of physics beyond the SM, whose 
scale $M$ is likely to be much higher than the electroweak scale. 
When $M \gg M_Z$, the effects of new physics at the high energy scale   
can be parametrized in all generality by adding a tower of {\it non-renormalizable}
effective operators to the SM Lagrangian 
which are $SU(3)\times SU(2)\times U(1)$ gauge invariant.
The coefficients of the higher-dimension operators are suppressed by inverse powers of 
the large energy scale $M$. 

The lowest dimension non-renormalizable operator is 
the well-known dimension-five $\Delta L=-2$ 
operator~\cite{weinberg} responsible for neutrino masses,
\bea
\label{O5}
{\cal {\delta L}}^{d=5} &\equiv& \frac{1}{2}
{\left(c_{5}\right)}_{\alpha \beta}\ O_{\alpha \beta}^{d=5} + \mrm{h.c.},\nn\\
O_{\alpha \beta}^{d=5} &=& \left( \overline{{\ell_L}^c_\alpha}\,  
{\widetilde\phi}^* \right) \, 
 \left( {\widetilde\phi}^\dag \, {\ell_L}_\beta \right) \ ,
\eea
where ${\ell_L}_\alpha$ are the lepton weak doublets with flavor denoted by the Greek index, the superscript $c$ denotes 
charge conjugation\footnote { The charge conjugate of the chiral fermion field 
is defined by ${\psi}^c \equiv \,i\gamma_0\gamma_2\, \overline{\psi}^T$.} 
 and $\tilde \phi$ is related to
the standard Higgs doublet by $\tilde \phi = i \tau_{2} \phi^*$. 
The precise value of $c_{5}$ is model-dependent.
The leading candidate high-energy theory of neutrino mass is the seesaw 
model \cite{seesaw}, the minimal extension of the SM which includes 
right-handed neutrino singlets $N_R$ with heavy Majorana masses.  The seesaw Lagrangian is 
\bea\label{Lag}
{\cal L}_{\rm seesaw} = &=& i \, \overline{\ell_L} \,
\D \, \ell_L + i \, \overline{e_R} \, \D \, e_R
+ i \, \overline{N_R} \, \dv \, N_R \nn\\
&&-  \overline{\ell_L} \,{\phi} \, {Y_e}  \, e_R
- \overline{\ell_L} \,{\widetilde\phi} \, {Y_\nu}  \, N_R
-\frac{1}{2}\,\overline{{N_R}^c} \, M \,{N_R}
+ \mrm{h.c.} \ , 
\eea
where $M$ denotes the Majorana mass matrix of the heavy neutrinos; and $Y_e$ and $Y_\nu$
are the Yukawa coupling matrices of the lepton doublets to the right-handed charged leptons and neutrinos, respectively.
In the seesaw model, the $d=5$ coefficient at the seesaw scale is given by
\be\label{cd5}
c_{5} = Y_\nu^* \, (M^*)^{-1} \, Y_\nu^\dagger \ .
\ee 
Upon spontaneous breaking of the electroweak gauge symmetry (EWSB), this irrelevant $d=5$ operator yields a relevant $d=3$ Majorana mass term for the
left-handed weakly interacting neutrinos.  The effective Majorana mass matrix of the light
neutrinos is given by 
\be
\left(m_\nu\right)_{\alpha \beta} \equiv -{v^2 \over {2}} \,\left( c_{5} \right)_{\alpha \beta}
\label{numass}
\ee
in the flavor basis.
Diagonalization of the effective mass matrix results in flavor-changing couplings of $W$ bosons to the leptonic charged currents. 

The RG evolution of  the $d=5$ operator coefficient $c_{5}$ of the unbroken theory 
is given by~\cite{d5RGE},
\bea
\label{RGE_c5}
16 \pi^2 \,  \frac{d}{d\ln\mu}c_{5}  &=& -\frac{3}{2} \left[\,c_{5}\,(Y_e {Y_e}^\dagger)
+(Y_e Y_e^\dagger)^T \,c_{5}\,\right]    \nonumber\\
&+&\left[-3\,g^2+2\,\lambda_H+ 2 \,{\rm Tr}\,\left(3 \,Y_u {Y_u}^\dagger + 3\, Y_d {Y_d}^\dagger + 
Y_e {Y_e}^\dagger\right)\right]\,c_{5} ,
\eea
where $\lambda_H$ is defined as the coupling of the Higgs potential term $-\lambda_H \left( \phi^\dagger \phi \right)^2/4$ in the SM Lagrangian.
At order $1/M$, the matrix $c_{5}$ is the only source of lepton flavor mixing, and 
it can be diagonalized by a unitary matrix $V$, whose evolution has been studied \cite{Casas-Espi}. 
Since $V$ is also the matrix which diagonalizes the light Majorana neutrino mass matrix in the 
 spontaneously broken theory at order $1/M$,
Eq.~(\ref{numass}),  it can be identified with the lepton
mixing matrix at this order.
For instance, for the simple case of two light lepton generations in the absence of CP-violation, the lepton mixing matrix is parametrized by a single 
angle $\theta$ whose RG evolution can be expressed 
in terms of the parameters of the broken theory. Eq.~(\ref{numass}) yields the RGEs
\bea
\label{RGE_d5}
\frac{d}{d\ln\mu} \cos\theta &=&- 
\frac{g^2}{32\pi^2 M_W^2}\,\frac{3}{2} \,\sin^2\theta\cos\theta\,
\frac{(m_{1}+m_{2})^2}{m_{2}^2-m_{1}^2}\,\left(m_{\mu}^2-m_{e}^2\right),
\nonumber \\
\frac{d}{d\ln\mu} \sin\theta &=& 
\frac{g^2}{32\pi^2 M_W^2}\,\frac{3}{2}\,\sin\theta\cos^2\theta\,
\frac{(m_{1}+m_{2})^2}{m_{2}^2-m_{1}^2}\,\left(m_{\mu}^2-m_{e}^2\right)\,,
\,
\eea
where $m_1$ and $m_2$ denote the two eigenvalues of the light neutrino Majorana mass matrix Eq.~(\ref{numass}).

The RGE of the lepton mixing matrix also has been obtained through the study of
flavor-changing one-loop corrections to low-energy observables. Ref.~\cite{PK} 
analyzes the case of the seesaw Lagrangian with both  heavy and light  Majorana neutrinos. 
In that case, the leptonic mixing matrix couples the charged leptons to both light and heavy neutrinos~\cite{SV}. In consequence, the matrix which couples exclusively to the light neutrinos is not unitary. 
The one-loop counterterms for that matrix result in RGE with a rich 
structure which goes beyond the approximations involved in obtaining \eq{RGE_d5}.
For two light neutrino generations in the absence of $CP$ violation, the RGE contains contributions not present in \eq{RGE_c5} above, proportional to
\bea
\frac{m_{e}^2+m_{\mu}^2}{m_{e}^2-m_{\mu}^2}\,\left(m_{1}^2-m_{2}^2\right)\,,
\label{excess}
\eea
as well as other terms present due to the non-unitarity of the mixing matrix.
A term with mass dependence similar to \eq{excess}
also appears in the RGE of the CKM quark mixing matrix \cite{Denner-Sack}. 

In this paper, we study the renormalization of the leptonic mixing matrix for the light Majorana neutrinos using the low-energy effective theory beyond the ${\cal O}(1/M)$ approximation of Eq.~(\ref{RGE_d5}).  
A term with the mass dependence shown in \eq{excess} is quadratic in the light neutrino masses $m_i$ and thus is 
an ${\cal O}(1/M^2)$ effect from the point of view of the effective theory. 
A proper comparison with the full theory requires the derivation of
the RGE in the effective theory to $O(1/M^2)$,  which includes operators of dimension $d=6$. 
The complete derivation of the one-loop RGE to $O(1/M^2)$ involves corrections from one-loop  diagrams containing 
two insertions of the $d=5$ operator or a single insertion of a $d=6$ operator.
In this work, we restrict ourselves to studying the dominant flavor-changing effects at ${\cal O}(1/M^2)$.
The Yukawa couplings $Y_\nu$ are
assumed to take natural values of ${\cal O}(1)$ throughout.  All corrections at order ${\cal O}(1/M)$ are included, but at order ${\cal O}(1/M^2)$ corrections suppressed by $Y_e^2$ and gauge couplings are neglected.
We also determine the relationship of these results to the counterterms obtained in the full theory in 
which both light and heavy neutrino species are present.

One could ask why a similar analysis is not necessary for the neutrino 
oscillation probability, and more concretely for its $\Delta m^2$ dependence, which is an ${\cal O}(1/M^2)$ effect. The answer is that the dominant one-loop corrections to the neutrino oscillation probability are those affecting the neutrino propagator, and while terms quadratic in $c_{5}$ give contributions to it in the seesaw model, terms linear in the $d=6$ operator coefficients do not.
In contrast, 
the definition at one loop of other low-energy observables, such as the 
$W$ decay width into leptons, does require inclusion 
of all $d=6$ operators, and their one-loop mixing with terms quadratic in the $d=5$ operator or linear in the $d=6$ operators.
In this sense, such low-energy observables depend on additional aspects of the new physics at high energy,
which are encoded in the $d=6$ operator coefficients of the low-energy effective theory.

In section 2, we obtain the RGEs for the effective theory before electroweak symmetry breaking (EWSB): the 
evolution of the $d=5$ operator is derived in the  $R_\xi$ gauge to ${\cal O}(1/M)$, and the RG evolution of the dominant flavor-changing 
contributions to ${\cal O}(1/M^2)$ is determined. Section 3 performs the same analysis after EWSB.
Section 4 compares the results for the effective theory with the counterterms obtained in the full theory in which heavy 
and light neutrino fields are present. Feynman rules and details of the computations are included in the 
Appendices.

\section{RGE before EWSB at ${\cal O}(1/M^2)$ }

The low-energy effective
Lagrangian of the minimal seesaw model to ${\cal O}(1/M^2)$ is of the form
\be\label{leff}
{\cal L}_{\rm eff} = {\cal L}_{\rm SM} + \delta{\cal L}^{d=5} + \delta{\cal
L}^{d=6} + \cdots,
\ee
where ${\cal L}_{\rm SM}$ is the SM Lagrangian and the 
higher-dimensional 
operators give the low-energy physics effects of the heavy Majorana neutrinos. 

At the seesaw scale, the $d=5$ term of the effective theory is given by Eqs.~(\ref{O5}) and~(\ref{cd5}). The $d=6$ term is~\cite{bgj}
\bea
\delta{\cal L}^{d=6} &=& (c_{6})_{\alpha \beta} \, O^{d=6}_{\alpha\beta}\,, \nn\\
\label{d6}
O^{d=6}_{\alpha\beta}&\equiv& \left( \overline{\ell_{L\alpha}} \widetilde \phi
\right)\, i \dv\, \left( \widetilde \phi^\dagger \ell_{L \beta} \right),
\eea 
where
\be\label{cd6}
c_{6} = Y_\nu \, (|M|^2)^{-1} \, Y_\nu^\dagger \ .
\ee  
When the Higgs doublet acquires a vacuum
expectation value, this irrelevant $d=6$ operator leads to a relevant $d=4$ flavor-nondiagonal kinetic energy term for
the left-handed Majorana neutrinos.  After normalization of the kinetic energy term and diagonalization of the kinetic energy and mass terms, the $d=6$ operator
ultimately results in a non-unitary 
correction to the leptonic charged currents and light Majorana neutrino neutral current~\cite{bgj}. 

At the seesaw scale, the coefficients of the operators $O^{d=5}$ and  $O^{d=6}$
are given by Eqs.~(\ref{cd5}) and (\ref{cd6}).
Additional $d=6$ operators will be present in the low-energy 
Lagrangian, however, since they are generated by radiative mixing in the renormalization
group running of the $d=6$ effective Lagrangian from the high-energy scale $M$ to the  
low-energy scale $\mu$:
\bea
\label{Ld6}
{\cal {\delta L}}^{d=6}(\mu) ={{\cal {\delta L}}^{d=6}(\mu)}^\dagger
= c_6(\mu)
\,O^{d=6} +\sum_i\left[ c_{i}(\mu)\ 
O_{i}(\mu) + {c_{i}(\mu)}^\dagger\ 
{O_{i}(\mu)}^\dagger \right] \,,
\eea
where $i$ runs over all independent $d=6$ operators.

For our purposes, we only need to consider those operators which are not flavor blind and contribute to the lepton 
masses/kinetic energies and  to the leptonic mixing matrix upon EWSB, that is, to those operators which ultimately modify either the charged and neutral currents
or the charged lepton Yukawa coupling matrix 
$Y_e$, up to ${\cal O}(1/M^2)$.

First, consider the one-loop corrections to the $d=5$ operator coefficient, the ${\cal O}(1/M)$ term. 
The complete one-loop renormalization of $c_5$ has been studied in the literature in 
the t'Hooft-Feynman gauge \cite{d5RGE}. We have re-derived the results in the 
covariant ${\rm R}_\xi$ gauge. The bare and renormalized $O_5$ coefficients are related by
\be
\label{c5B}
c_5^{\rm Bare} = \left( Z_\phi \right)^{-1/2} \left(Z_{\ell_L}^T \right)^{-1/2}
 \left[ c_5  + \delta c_5 \right]
\mu^{2\epsilon}\left( Z_{\ell_L}\right)^{-1/2} \left( Z_\phi \right)^{-1/2},
\ee
where $2\epsilon=4-d$, and $Z_\phi$ and $Z_{\ell_L}$ are the one-loop
wave-function renormalization constants of the scalar and lepton doublets in the $\overline{MS}$ minimal subtraction scheme:
\bea
\label{wf}
Z_{\phi}&=&1-\frac{1}{16\pi^2}\,\frac{1}{\epsilon}
             \bigg\{{\rm Tr}\,\left[Y_eY_e^{\dagger}+
              3Y_uY_u^{\dagger}+3Y_dY_d^{\dagger}\right]+\lambda_H\nn\\
              &&\qquad\qquad\qquad
              - \frac{3g^2}{2}\left[1-\frac{1}{2}(\xi-1)\right] 
              - \frac{(g')^2}{2}\left[1-\frac{1}{2}(\xi-1)\right] \bigg\}\,,\nn\\\\
Z_{\ell_L}&=&1-\frac{1}{16\pi^2}\,\frac{1}{\epsilon}\,
 \bigg\{\frac{Y_eY_e^{\dagger}}{2}
              + \frac{3g^2}{4}\left[1+(\xi-1)\right] 
              + \frac{(g')^2}{4}\left[1+(\xi-1)\right] \bigg\}\,.
\eea
The vertex counterterm $\delta c_5$ is
\bea
\label{delta_c5}
\delta c_5 =-\frac{1}{16 \pi^2}\, \frac{1}{\epsilon}\,
 \bigg\{c_5 \,(Y_eY_e^\dagger)+(Y_eY_e^\dagger)^T\, c_5 +
 \frac{3g^2}{2}\left[\frac{1}{2}+(\xi-1)\right]\,c_5
             -\frac{(g')^2}{2}\left[\frac{1}{2}-(\xi-1)\right]\,c_5  \bigg\}\,.
\eea
From Eqs.~(\ref{c5B})-(\ref{delta_c5}), one obtains the RGE of the $d=5$ operator given 
in \eq{RGE_c5}.

Now, consider ${\cal O}(1/M^2)$ corrections. 
The $d=6$ operators do not correct $O^{d=5}$ at ${\cal O}(1/M^2)$, as there are 
no $SU(3)\times SU(2)\times U(1)$ gauge invariant and $(B-L)$-non invariant $d=6$ operators~\cite{bgj}.
The coefficients of $d=6$ operators are corrected at ${\cal O}(1/M^2)$ by effects quadratic in the $d=5$ operator and by effects linear in the $d=6$ operators.
We will restrict our analysis to ${\cal O}(v^2/M^2)$ and ${\cal O}(Y_e\ v^2/M^2)$ corrections, disregarding 
in this work ${\cal O}(g^2\ v^2/M^2)$,  ${\cal O}(g'^2\ v^2/M^2)$  and ${\cal O}(Y_e^2\ v^2/M^2)$ contributions.  There are no corrections proportional 
to the Higgs self-coupling $\lambda_H$ to ${\cal O}(v^2/M^2)$.

We classify the possible $d=6$ operators by using the operator basis defined in Ref.~\cite{BW}, with the following exception: the weak singlet and triplet operators  
\bea
\label{sing}
O^{(1)}_{\phi\ell}&=& \,\left( {\overline\ell_L}_\alpha \,\gamma_\mu\,{\ell_L}_\beta \right ) \;
\left(\,\phi^\dagger\,D^\mu\phi \right)\,,\\
\label{trip}
O^{(3)}_{\phi\ell}&=& 
(\overline{\ell_L}_\alpha \,\vec\tau\,\gamma_\mu\,{\ell_L}_\beta) \;
 \left({\phi}^\dagger\,\vec\tau\,D^\mu\phi\right)\,,
\eea
are replaced by the orthogonal combinations
\bea
\label{mp}
&& O^{(-)}_{\phi\ell}=\frac{1}{2}\left[ O^{(1)}_{\phi\ell}-O^{(3)}_{\phi\ell} \right] \equiv O^{d=6}\,,\nn\\
&&O^{(+)}_{\phi\ell}= \frac{1}{2}\left[ O^{(1)}_{\phi\ell}+O^{(3)}_{\phi\ell} \right]\,.
\eea
This replacement is advantageous because $O^{(-)}_{\phi\ell}$ is equal to the $d=6$ operator $O^{d=6}$, \eq{d6}.

The one-loop diagrams resulting in relevant ${\cal O}(1/M^2)$ corrections from two insertions of the 
$O^{d=5}$ operator are given in Fig.~1.  These diagrams result in contributions to several 
operators defined in the basis of Ref.~\cite{BW}. Details of the computation of the diagrams in the Feynman 
gauge are given in Appendix \ref{appendixB}.
The contributions relevant to the flavor-mixing renormalization
program modify the coefficients $c_{\phi\ell}^{(-)}\equiv c_6$ and
$c_{\phi\ell}^{(+)}$ corresponding to the operators in \eq{mp}, and
the coefficient $c_{e\phi}$ corresponding to the operator
\be
\label{O_ephi}
O_{e\phi}= (\overline\ell_L\,\phi\,e_R)\,\left(\, {\phi}^\dagger\phi \right)\, ,
\ee
defined in Ref.~\cite{BW}.  As will be discussed in the next section, 
$O_{e\phi}$ is important for our analysis since it results in flavor non-diagonal corrections to  the charged lepton masses,
and thus to flavor mixing in charged currents when the Higgs acquires a vacuum expectation value.
$O^{(+)}_{\phi\ell}$ also is kept as it contributes to flavor-changing charged and neutral currents.

The one-loop diagrams (see Appendix \ref{appendixB}) result in the counterterms:
\bea
\label{dcct}
&&\delta c_{\phi \ell}^{(-)}\equiv \delta c_6=  \frac{1}{16\pi^2}\, \frac{1}{\epsilon} \; \frac{9}{4} \,({c_{5}}^*c_{5})\,,\\
&&\delta c^{(+)}_{\phi\ell}=  \frac{1}{16\pi^2}\, \frac{1}{\epsilon} \; \frac{1}{4}\,({c_{5}}^*c_{5})\,,\\
&&\delta c_{e\phi}= -\frac{1}{16\pi^2}\, \frac{1}{\epsilon} \;\frac{3}{4}\,
\left({c_{5}}^*c_{5}\;Y_e\right)\, ,
\eea
where the counterterms are defined to remove the divergent $1/\epsilon$ poles of the
one-loop graphs in the $\overline{MS}$ scheme.

\begin{picture}(300,100)(-50,0)
\label{d1a}
\ArrowLine(0,0)(40,40)
\ArrowLine(10,20)(20,30)
\put(-20,10){${\ell_{L \,\beta}}$}
\ArrowLine(80,40)(120,0)
\ArrowLine(100,30)(110,20)
\put(125,10){${\ell_{L \,\alpha}}$}
\DashArrowLine(0,80)(40,40){5}
\put(-20,70){ $\phi$}
\DashArrowLine(80,40)(120,80){5}
\put(125,70){ $\phi$}
\BBoxc(40,40)(6,6)
\BBoxc(80,40)(6,6)
\DashArrowArc(60,40)(17,0,180){5}
\put(50,10){ ${\ell_{L\,\gamma}}$}
\ArrowLine(77,40)(43,40)
\ArrowLine(53,30)(67,30)
\put(50,67){ $\phi_l$}
\DashArrowLine(200,0)(240,40){10}
\put(185,10){$\phi$}
\DashArrowLine(280,40)(320,0){10}
\put(320,10){$\phi\,$}
\DashArrowLine(200,80)(240,40){10}
\put(185,70){ $\phi$}
\DashArrowLine(280,40)(320,80){10}
\put(320,70){ $\phi$}
\BBoxc(240,40)(6,6)
\BBoxc(280,40)(6,6)
\ArrowArc(260,40)(17,0,180)
\ArrowArc(260,40)(27,75,105)
\put(257,77){${\ell_{L\,\alpha}}$}
\ArrowLine(277,40)(243,40)
\ArrowLine(253,30)(267,30)
\put(250,10){ ${\ell_{L \,\beta}}$}
\end{picture}
\vspace{1cm}

\begin{picture}(300,100)(-50,0)
\ArrowLine(0,0)(40,40)
\ArrowLine(10,20)(20,30)
\put(-20,17){${\ell_{L\,\delta}}_d$}
\ArrowLine(80,40)(120,0)
\ArrowLine(20,50)(10,60)
\put(120,17){${\ell_{L \,\beta}}_b$}
\ArrowLine(0,80)(40,40)
\ArrowLine(20,50)(10,60)
\put(-20,60){ ${\ell_{L\,\gamma}}_c$}
\ArrowLine(80,40)(120,80)
\ArrowLine(100,30)(110,20)
\ArrowLine(110,60)(100,50)
\put(120,60){ ${\ell_{L\,\alpha}}_a$}
\DashArrowArc(60,40)(17,0,180){5}
\put(50,20){ $\phi_n$}
\DashArrowLine(77,40)(43,40){5}
\put(50,67){ $\phi_m$}
\BBoxc(40,40)(6,6)
\BBoxc(80,40)(6,6)
\ArrowLine(210,0)(240,30)
\put(190,5){ ${e_{R\,\beta}}$}
\ArrowLine(280,60)(320,10)
\ArrowLine(300,50)(305,42)
\put(315,25){ ${\ell_{L \,\alpha}}$}
\DashArrowLine(200,95)(240,60){5}
\put(190,75){ $\phi$}
\DashArrowLine(230,105)(240,60){5}
\put(230,110){ $\phi$}
\DashArrowLine(280,60)(320,100){5}
\put(320,90){ $\phi$}
\ArrowLine(240,30)(240,60)
\ArrowLine(230,35)(230,45)
\put(195,40){ ${\ell_{L\,\gamma}}$}
\ArrowLine(280,60)(240,60)
\ArrowLine(255,70)(265,70)
\put(250,80){ ${\ell_{L \,\delta}}$}
\DashArrowLine(280,60)(240,30){5}
\put(260,30){ $\phi$}
\BBoxc(240,60)(6,6)
\BBoxc(280,60)(6,6)
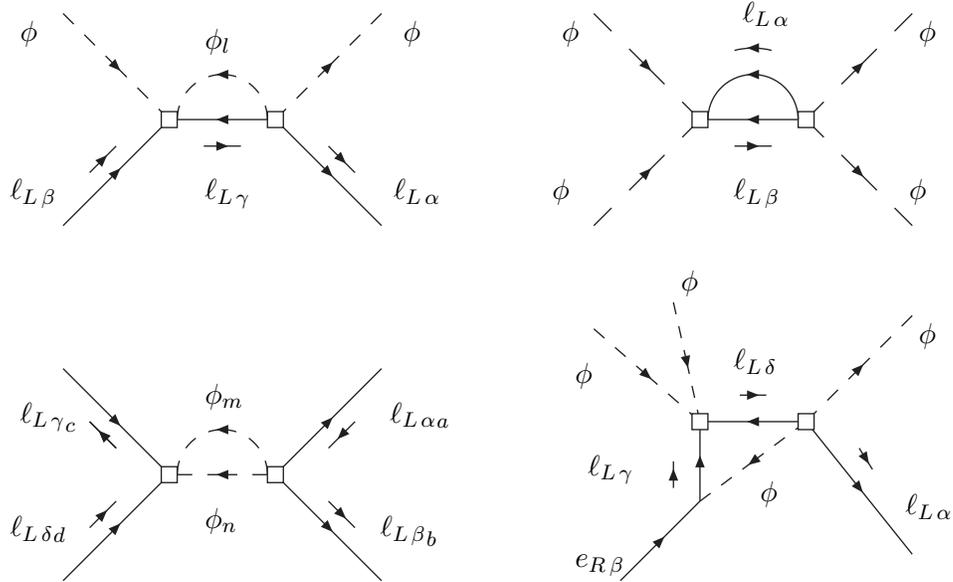
\put(-40,-20){ {\rm \scriptsize{\bf{Figure 1}: One-loop diagrams with
two insertions of the $d=5$ operator. The extra arrows next  }}} 
\put(-40,-30){ {\rm \scriptsize{\bf{
to lepton indices
indicate fermion flow, as defined in Ref. \cite{Denner}}}}}
\end{picture}
\vspace{2cm}

The bare and renormalized coefficients of the operators  $O^{(-)}_{\phi\ell}\equiv O^{d=6}$, $O^{(+)}_{\phi\ell}$ and  $O_{e\phi}$
are related by
\bea
&&{c^{(\pm)}_{\phi\ell}}^{\rm Bare} =\left(Z_\phi^\dagger \right)^{-1/2} \left( Z_{\ell_L}^\dagger \right)^{-1/2}
 \left[ c^{(\pm)}_{\phi\ell}+ \delta  c^{(\pm)}_{\phi\ell} \right]\,  \mu^{2\epsilon}\,
\left( Z_{\ell_L}\right)^{-1/2} \left( Z_\phi\right)^{-1/2}\,,\\
 \label{cepbare}
&&c_{e\phi}^{\rm Bare}= \left(Z_\phi^\dagger\right)^{-1/2} \, \left(Z_{\phi}\right)^{-1/2}\, \;\left(Z_{\ell_L}^\dagger\right)^{-1/2}
\left[ c_{e\phi}+ \delta c_{e\phi} \right] \, \mu^{2\epsilon} \,
\left(Z_\phi\right)^{-1/2}\,\left( Z_{e_R}\right)^{-1/2}\,,
\eea
Eqs.~(\ref{dcct})-(\ref{cepbare}) lead finally to the RGEs:
\bea\label{RGEc6}
&&\frac{d }{d\ln \mu} c^{(-)}_{\phi\ell} \equiv  \frac{d}{d\ln\mu}c_6= \frac{1}{16\pi^2}\,\frac{9}{2}\, (c_5^*c_5)\,,\\
&&\frac{d }{d\ln \mu}  c^{(+)}_{\phi\ell}= \frac{1}{16\pi^2}\,\frac{1}{2}\, (c_5^*c_5)\,,\label{RGEcplus}\\
&&\frac{d}{d \ln\mu} c_{e\phi}= - \frac{1}{16\pi^2}\,\frac{3}{2}\,\left({c_{5}}^*c_{5}\;Y_e\right)\ .\label{RGEcephi}
\eea

These are the results before EWSB. It is necessary to consider the spontaneously broken theory in order to compare with low-energy experiments.

\section{RGE after EWSB at ${\cal O}(1/M^2)$ }
There is an important difference between the effective mixing matrix defined 
at tree-level and at one-loop, when EWSB is taken into account.
At tree-level to ${\cal O}(1/M^2)$, all charged and neutral-current mixing effects stem from 
flavor non-diagonal terms in the neutrino masses and neutrino kinetic energies, which arise from
the operator coefficients $c_5$ and $c_6$ of the effective theory.
At one-loop to ${\cal O}(1/M^2)$, however, additional effects arise from flavor non-diagonal contributions to the charged lepton masses.

\subsection{Tree-level effective Lagrangian}

After spontaneous electroweak symmetry breaking, the relevant part of the $d\le 6$ effective Lagrangian is given by 
\bea
\label{Leff}
{\cal L}_{\rm eff}
&=&i\,\overline e_{L \alpha}\,\dv \;e_{L \alpha}+
i\,\overline e_{R \alpha}\,\dv\;e_{R \alpha}
-\left(\overline{e_L}_\alpha \, (m_e)_{\alpha \beta}
\,{e_R}_\beta + \mrm{h.c.} \right) \nn\\
&+&
i\,\overline\nu_{L \alpha}\,\dv \, \left( \delta_{\alpha \beta} + {\lambda_{\alpha \beta}} \right)
\, \nu_{L \beta}
-\frac{1}{2}\,
\left( \overline{{\nu_L}^c}_\alpha\,(m_\nu)_{\alpha\beta}\,{\nu_L}_\beta
+ \mrm{h.c.} \right) \\
&+& {\cal L}_{\rm W,Z} + {\cal L}_{\rm Higgs}\,,\nn 
\eea
where 
\be
{\cal L}_{\rm W,Z}= \frac{g}{\sqrt{2}}
\left(J_\mu^{-\, CC}\, W^{+ \mu}
+ J_\mu^{+\, CC}\, W^{- \mu} \right)
+\frac{g}{\cos\theta_W}J_\mu^{NC}\, Z^\mu
\ee
contains the $W$ and $Z$ couplings to 
the weak charged and neutral currents, and
${\cal L}_{\rm Higgs}$ contains the lepton couplings to the
Higgs boson.

In the flavor basis, the charged and neutral currents are given by
\bea
J_\mu^{-\, CC}\,&=& \overline e_{L \alpha} \gamma_\mu \nu_{L \alpha}\,,\\
J_\mu^{NC}\,&=& {1 \over 2} \overline \nu_{L \alpha} \gamma_\mu
\nu_{L \alpha} + \left({-{1 \over 2} -s^2\theta_W }\right)
\overline e_{L \alpha} \gamma_\mu e_{L \alpha}
+ \left(-s^2\theta_W \right) \overline e_{R \alpha} \gamma_\mu e_{R \alpha}\,.\nn
\eea

The charged lepton mass matrix, the light Majorana neutrino mass matrix and the 
neutrino kinetic energy flavor-changing effective coupling at tree-level  are given by
\bea
m_e &\equiv & \frac{v}{\sqrt{2}}\,Y_e \,,\nn\\
m_\nu &\equiv & -{v^2 \over {2}} \,c_{5} \,, \\
\lambda &\equiv& {v^2 \over {2}} \,c_{\phi\ell}^{(-)}= {v^2 \over {2}} \,c_{6}. \nn
\eea

As shown in Ref.~\cite{bgj},
the effect of the $d=6$ operator coefficient $\lambda$ is transferred to the weak currents by redefining the neutrino field to the canonical basis after EWSB (or by using the equations of motion of $\ell_L$ before EWSB).  The net effect is a rescaling of the neutrino field $\nu_{L \alpha} \to ( {\mathbb I}- \frac \lambda 2 )_{\alpha \beta} \nu_{L \beta}$ in the effective Lagrangian Eq.~(\ref{Leff}). After the rescaling, the charged and neutral currents are given by
\bea
&&J_\mu^{-\, CC}\,= \overline e_{L \alpha} \gamma_\mu 
\left(\delta_{\alpha\beta}-\frac{(\lambda_{CC})_{\alpha\beta}}{2}\right)\,\nu_{L \beta}\,,\\
&&J_\mu^{NC}\,= {1 \over 2} \overline \nu_{L \alpha} \gamma_\mu\,
\left(\delta_{\alpha\beta}-(\lambda_{NC}^\nu)_{\alpha\beta}\right)\,\nu_{L \beta}\,
+ \left({-{1 \over 2} -s^2\theta_W }\right)
\overline e_{L \alpha} \gamma_\mu e_{L \alpha}
+ \left(-s^2\theta_W \right) \overline e_{R \alpha} \gamma_\mu e_{R \alpha}\,,\nn
\eea
where
\be
\lambda_{CC}=\lambda_{NC}^\nu=\lambda
\ee
at tree-level. We will see in the following subsection that
$\lambda_{CC}$ and $\lambda_{NC}^\nu$ have different one-loop corrections.

It is customary to define the tree-level matrices in the flavor basis where
$m_e$ is diagonal, but $m_\nu$ and $\lambda$ are not.  The mass basis is obtained from the
flavor basis by making a unitary transformation $V$ which diagonalizes the light Majorana neutrino mass matrix,
\be
\label{diag_nu}
 V^* \, m_\nu \, V^\dagger =\left( {m_\nu}\right)_{\rm diag, real}\,.
\ee
The light neutrino Majorana mass eigenstates $\nu_i = \nu_i^c$,
$i=1,\cdots, n$, are defined by 
\be
\nu_i= V_{i \alpha } \,{\nu_L}_\alpha
+ V_{i \alpha }^*\,{\nu_L}_\alpha^c ~,
\ee
so
\be
{\nu_L}_\alpha = P_L \, \left( V^\dagger \right)_{\alpha i} \, \nu_i  = P_L \, V^*_{i \alpha} \, \nu_i \ .
\ee

The weak currents in the mass eigenstate basis
\bea
\label{JCC_d6}
J_\mu^{-\,CC}  
&=&\overline {e_L}_\alpha \, \gamma_\mu \,
U^{\rm{eff}}_{\alpha i} \, \nu_i\,,\\\nn
\label{JNC_d6}
J_\mu^{NC}&=& {1 \over 2} \,\overline \nu_i \,\gamma_\mu
\left( {U_{NC}^{\rm eff}}^\dagger  {U_{NC}^{\rm eff}} \right) _{i j}
\,\nu_j + \left({-{1 \over 2} -s^2\theta_W }\right)
\overline e_{L \alpha} \gamma_\mu e_{L \alpha}
+ \left(-s^2\theta_W \right) \overline e_{R \alpha} \gamma_\mu e_{R \alpha}\,\nn
\eea
are defined in terms of the non-unitary effective leptonic mixing matrices
\bea
\label{U_eff_CC}
&&U^{\rm{eff}}_{CC} \equiv
\left({\mathbb I}-\frac{\lambda_{CC}}{2}\right)\, V^\dagger \, , \\
\label{U_eff_NC}
&&U^{\rm{eff}}_{NC}\equiv
\left({\mathbb I}-\frac{\lambda^\nu_{NC}}{2}\right)\, V^\dagger \, .
\eea
At tree-level, $ U^{\rm{eff}}_{CC} = U^{\rm{eff}}_{NC} $. 

Recall that $m_\nu$ does not get a contribution from $\lambda$ up to order ${\cal O}(1/M^{3})$.
(The rephasing $\Omega$ of the charged lepton fields in Ref.~\cite{bgj} is ignored here.).

\subsection{One-loop effective Lagrangian}

The one-loop effective Lagrangian is given by
\be
{\cal L}_{\rm eff}^{\rm Bare}= 
{\cal L}_{\rm eff} +
\delta {\cal L}_{\rm eff} 
\ee
where ${\cal L}_{\rm eff}$ is the renormalized Lagrangian and $\delta{\cal L}_{\rm eff}$ is the counterterm Lagrangian which cancels the ultraviolet (UV) divergences of the one-loop diagrams.  In the canonical flavor basis, the counterterms relevant for one-loop leptonic mixing are
\bea
\delta{\cal L}_{\rm eff}
&=&i\, \left( \delta Z_{e_L} \right)_{\alpha \beta}\, \overline e_{L \alpha}\,\dv \;e_{L \beta}+
i\, \left( \delta Z_{e_R} \right)_{\alpha \beta} \,\overline e_{R \alpha}\,\dv\;e_{R \beta}
-\left(\overline{e_L}_\alpha \, (\delta m_e)_{\alpha \beta}
\,{e_R}_\beta + \mrm{h.c.} \right) \nn\\
&+&
i\,\left( \delta Z_{\nu_L} \right)_{\alpha \beta}\,
\overline\nu_{L \alpha}\,\dv \,\, \nu_{L \beta} 
\,-\frac{1}{2}\,
\left( \overline{{\nu_L}^c}_\alpha\,(\delta m_\nu)_{\alpha\beta}\,{\nu_L}_\beta
+ \mrm{h.c.} \right) \\
&+& 
\frac{g}{\sqrt{2}}\,\left(
\overline e_{L\alpha} \, \gamma_\mu \,(\delta\lambda_{CC})_{\alpha \beta}
\nu_{L \beta}\,W^{+ \mu} + \mrm{h.c.}\right)\nn\\
&+& 
\frac{g}{\cos\theta_W} Z^\mu 
\left[{1 \over 2} \overline \nu_{L \alpha} \gamma_\mu\, (\delta\lambda^\nu_{NC})_{\alpha \beta} \,\nu_{L \beta} 
-{1 \over 2}
\overline e_{L \alpha} \gamma_\mu  \,(\delta\lambda_{NC}^e)_{\alpha \beta}\,e_{L \beta} \right] .
\nn
\eea
Coupling constant counterterms $\delta g$, $\delta \theta_W$
are not relevant for the mixing analysis and have been omitted. 

In the $\overline{MS}$ scheme, the bare and renormalized quantities are related by
\bea
&&m_e^{\rm Bare}=
\;(Z_{e_L}^\dagger)^{-1/2} \left[ m_{e}+ \delta m_{e} \right] \, \mu^{2\epsilon} \,
Z_{e_R}^{-1/2}\,,\\
&& m_\nu^{\rm Bare} =  \left(Z_{\nu_L}^T \right)^{-1/2}
 \left[  m_\nu  + \delta  m_\nu \right]
\mu^{2\epsilon}\, Z_{\nu_L}^{-1/2}\,,\\
&&
\lambda_{CC}^{\rm Bare} =  \left(Z_{e_L}^\dag \right)^{-1/2}
 \left[  \lambda_{CC}  + \delta  \lambda_{CC} \right]
\, Z_{\nu_L}^{-1/2} \,,\\
&&
(\lambda_{NC}^\nu)^{\rm Bare} =  \left(Z_{\nu_L}^\dag \right)^{-1/2}
 \left[  \lambda^\nu_{NC}  + \delta  \lambda^\nu_{NC} \right]
\, Z_{\nu_L}^{-1/2} \,,\\
&&
(\lambda_{NC}^e)^{\rm Bare} =  \left(Z_{e_L}^\dag \right)^{-1/2}
 \left[  \lambda^e_{NC}  + \delta  \lambda^e_{NC} \right]
\, Z_{e_L}^{-1/2} \,.
\label{ren_m}
\eea

In general, the one-loop wavefunction renormalization constants $Z_{e_L}$ and $Z_{\nu_L}$ are modified from $Z_{\ell_L}$ in \eq{wf} by the diagrams in Fig.~2:
\bea
Z_{e_L} &=& Z_{\ell_L}-\frac{1}{16\pi^2}\,\frac{1}{\epsilon}\,M_W^2 \,c^{(-)}_{\phi\ell} ,
\label{wf_e}
\\
Z_{\nu_L} &=& Z_{\ell_L}-\frac{1}{16\pi^2}\,\frac{1}{\epsilon}\, \left(M_Z^2+M_H^2\right) \,c^{(-)}_{\phi\ell} .
\label{wf_nu}
\eea
\begin{picture}(200,120)(0,0)
\ArrowLine(60,40)(120,40)
\put(80,30){${e_{L_\beta}}$}
\ArrowLine(120,40)(180,40)
\put(140,30){${e_{L_\alpha}}$}
\DashArrowArc(120,70)(30,-90,270){5}
\put(110,80){ $\phi^+_l$}
\BCirc(120,40){3}
\ArrowLine(260,40)(320,40)
\put(280,30){${\nu_{L_\beta}}$}
\ArrowLine(320,40)(380,40)
\put(340,30){${\nu_{L_\alpha}}$}
\DashArrowArc(320,70)(30,-90,270){5}
\put(310,80){ $\phi^{0}_l$}
\BCirc(320,40){3}
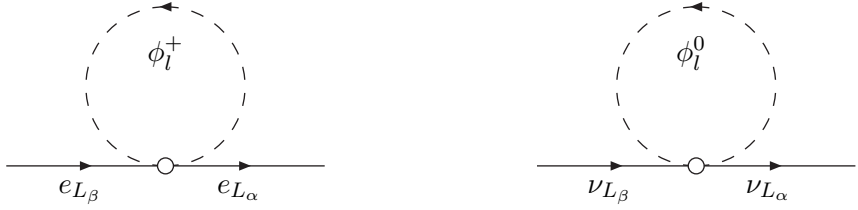
\put(60,10){ {\rm \scriptsize{\bf{Figure 2}: One-loop diagrams with
one insertion of the $d=6$ operator.}}} 
\end{picture}

In our analysis, we neglect these additional contributions since they are ${\cal O}(g^2 v^2/M^2)$. For completeness, we also give here the wavefunction renormalization constant of the right-handed charged lepton in the $R_\xi$ gauge:
\be
Z_{e_R} = 1-\frac{1}{16\pi^2}\,\frac{1}{\epsilon}
\bigg\{\frac{Y_e^\dag Y_e}{2}+g'^2[1+(\xi-1)]\bigg\}\ .
\ee

The vertex counterterms are given by 
\bea
\label{delta_m}
&&\delta m_e = \frac{v}{\sqrt{2}} \, \delta Y_e + \left(\frac{v}{\sqrt{2}}\right)^{3} \delta  c_{e\phi}+  \frac{\delta v}{\sqrt{2}}  \, Y_e\,, \\
&&\delta m_\nu = -\frac{v^2}{2} \delta c_5 - \frac{\delta v^2}{2} c_5\,, \\
&&\delta \lambda_{CC} = \frac{v^2}{2}\,
\frac{ \delta c^{(-)}_{\phi\ell}-\delta  c^{(+)}_{\phi\ell}}{2}
 \,,\\
&&\delta \lambda^\nu_{NC} = \frac{v^2}{2}\,
\delta  c^{(-)}_{\phi\ell} \,,\\
&&\delta \lambda^e_{NC} = \frac{v^2}{2}\,
\delta  c^{(+)}_{\phi\ell} \,.\eea

Notice that flavor-changing corrections to the left-handed electron neutral current are generated at one-loop.
\subsection{RGE of $U^{\rm eff}_{CC}$}

The operator coefficient $\delta c_{e\phi}$ 
induces flavor-nondiagonal one-loop corrections to the charged lepton mass matrix, so it is necessary to diagonalize the charged lepton mass matrix at one-loop.  The unitary matrix $E$ which diagonalizes the charged lepton mass matrix is defined by 
\bea
\label{diag_e}
&&E\, (m_e m_e^\dagger ) \, E^\dagger 
\equiv (m_e^2)_{\rm diag, real}\,.
\eea
The relation of $E$ to its bare matrix is
\be
E^{\rm Bare}= \mathbb I= E +\delta E \ .
\ee
The presence of a non-trivial matrix $E$ changes the definition of the charged-current lepton 
mixing matrix, 
\bea
&&J_\mu^{-\,CC}  
=\overline {e_L}_\alpha \, \gamma_\mu \,
({{U^{\rm{eff}}_{CC}}})_{\alpha i} \, \nu_i\,,
\eea
which is defined at one-loop by the renormalized non-unitary leptonic mixing matrix
\be
U^{\rm eff}_{CC} \equiv E\left({\mathbb I}-\frac{\lambda_{CC}}{2}\right)   V^\dag  \,.
\ee
Note that the product $EV^\dagger$ is the leptonic analogue of the CKM matrix. 

The RGE of $U^{\rm eff}_{CC}$ is then given by
\be
\label{RGE_Ueff}
\frac{d {U^{\rm{eff}}_{CC}}}{d\ln\mu}  =  T_e\,{U^{\rm{eff}}_{CC}} - {U^{\rm{eff}}_{CC}}\,T_\nu\,
-\frac{1}{2} E\,\frac{d \lambda_{CC}} {d\ln\mu} V^\dag\, ,
\ee
where $T_e$ and $T_\nu$ are anti-Hermitian auxiliary matrices introduced to maintain the unitarity 
of $E$ and $V$ under renormalization group running,
\bea\label{RGEV}
\frac{d E}{d\ln\mu}  &\equiv& T_e\,E\,, \nn\\
\frac{d V}{d\ln\mu}  &\equiv&  T_\nu\,V\,.
\eea
$T_e$ can be derived from \eq{diag_e},
\bea
\label{dme}
 {{d} \over {d\ln\mu}} (m_e^2)_{\rm diag, real} &=&  T_e\, (m_e^2)_{\rm diag, real}-(m_e^2)_{\rm diag, real}\,T_e +  E \left[{{d m_e} \over {d\ln\mu}} m_e^\dag + m_e{{d m_e^\dag} \over {d\ln\mu}} \right]\,E^\dag \ ,\nn\\
\eea
with the RGE for the charged lepton mass matrix given by
\bea
{{d m_e} \over {d\ln\mu}}&=& 
\frac{v}{\sqrt{2}} \, \frac{d Y_e}{d\ln\mu}  +  \frac{1}{\sqrt{2}}\,
\frac{d v}{d\ln\mu}\,  Y_e\, + \left(\frac{v}{\sqrt{2}}\right)^{3}
\frac{d c_{e\phi}}{d\ln\mu} 
-\frac{1}{2}\frac{d Z_{e_L}^\dag}{d \ln\mu}\,m_e\,
-m_e\,\frac{1}{2}\frac{d Z_{e_R}}{d \ln\mu}\,. 
\eea

Imposing the constraint that the off-diagonal matrix elements on the RHS of \eq{dme} vanish to order ${\cal O}(1/M^2)$, it follows that
\bea
&&(T_e)_{\alpha\beta}=
\Bigg\{\begin{array}{cc} 
0& \qquad,\ \alpha=\beta\\
-\frac{1}{16\pi^2}\,\frac{g^2}{M_W^2}\,\frac{m_\alpha^2+m_\beta^2}{m_\alpha^2-m_\beta^2}\,
\frac{3}{2} \,\sum_i\,\,V_{i \alpha}^*\,|{m_\nu}_i|^2 \,V_{i\beta }
& \qquad ,\, \alpha\neq \beta\\
\label{Te}
\end{array}
\eea
where the non-diagonal terms stem from the RGE of the operator coefficient $c_{e\phi}$.

A similar expression can be found for $T_\nu$ \cite{Chankowski}:
\bea
&&Re\left[(T_\nu)_{ij}\right]=
\Bigg\{\begin{array}{cc} 
0& \qquad i=j\\
-\frac{1}{16\pi^2}\,\frac{3}{2}\,\frac{g^2}{2M_W^2}\,
\frac{m_i+m_j}{m_i-m_j}\,
Re\left[\sum_\alpha\,V_{i\alpha }\,|{m_e}_\alpha|^2\,V_{j\alpha }^*\right]
& \qquad i\neq j\\
\end{array}\\\nn\\
&&Im\left[(T_\nu)_{ij}\right]=
\Bigg\{\begin{array}{cc} 
0& \qquad i=j\\
-\frac{1}{16\pi^2}\frac{3}{2}\,\frac{g^2}{2M_W^2}\,
\frac{m_i-m_j}{m_i+m_j}\,
Im\left[\sum_\alpha\,V_{i\alpha }\,|{m_e}_\alpha|^2\,V_{j\alpha }^*\right]
& \qquad i\neq j\\
\end{array}
\label{Tnu}
\eea

Finally, the last term in \eq{RGE_Ueff} can be rewritten as
\be
\frac{1}{2} \,E\,\frac{d \lambda_{CC}} {d\ln\mu} V^\dag= 
\frac{v^2}{2}\,\frac{1}{4} 
\,E\,\left(\frac{d c_{\phi\ell}^{(-)}}{d\ln\mu}- \frac{d c_{\phi\ell}^{(+)}}{d\ln\mu}\right) V^\dag\,= 
\frac{1}{16\pi^2}\, \frac{g^2}{2\,M_W^2}\, V^\dagger \,(m_\nu)^2_{\rm diag,real}\, 
\ee
using Eqs.~(\ref{RGEc6}), (\ref{RGEcplus}) and~(\ref{delta_m}).

Thus, \eq{RGE_Ueff} becomes
\bea
\frac{d \left({U^{\rm{eff}}_{CC}} \right)_{\alpha i} }{d\ln\mu}  &=& 
\frac{g^2}{32 \pi^2 M_W^2}\,\Bigg\{
\,\frac{3}{2}\,\sum\limits^{n}_{\beta\neq \alpha}\,
\frac{ m^2_\beta+ m^2_{\alpha} }{m^2_\beta - m^2_{\alpha}}\, V^*_{j\alpha }\,m_j^2\,{V}_{j \beta }\,(U^{\rm{eff}}_{CC})_{\beta i}
 - V^*_{i \alpha}\,m_i^2\\
&&\hspace{1,3cm}+\frac{3}{2}\,\sum\limits_{j\neq i}^{n} 
\frac{(U_{CC}^{\rm{eff}})_{\alpha j}}{m^2_j - m^2_i}\,
\bigg[\,(m^2_j + m^2_i)\, 
{V_{j \beta }}\,
m^2_\beta\,V^*_{i \beta }\,
+\: 2m_j m_i\, 
V_{j\beta }^*\, m^2_\beta\,
V_{i\beta }\,
\bigg]\,\Bigg\}\ .\nn
\,
\eea

For illustrative purposes, let us write explicitly the result 
for two generations of light fermions in the absence of $CP$ violation. The effective mixing matrix at tree level is then parametrized by
\bea
 {U^{\rm{eff}}_{CC}}= 
\left(\begin{array}{cc}
1- \frac{\lambda_{ee}}{2}& - \frac{\lambda_{e \mu}}{2} \\
- \frac{\lambda_{e \mu}}{2}& 1- \frac{\lambda_{\mu\mu}}{2} \\
\end{array}\right)\,
\left(\begin{array}{cc} 
\cos\theta &\sin\theta\\
-\sin\theta &\cos\theta
\end{array}\right)\,\, .
\label{U_eff_2}
\eea
The corresponding RGE of the matrix elements $\left(U^{\rm eff}_{CC}\right)_{ei}$ are
\bea
\frac{d \left( {U^{\rm{eff}}_{CC}}\right)_{e1} }{d\ln\mu}  &=& 
-\frac{g^2}{32 \pi^2 M_W^2}\,\Bigg\{
\,\frac{3}{2}\,
\bigg[\frac{ m^2_{\mu}+ m^2_e }{m^2_{\mu} - m^2_e}\,
(m_2^2-m_1^2) +
\frac{ (m_2+ m_1)^2 }{m^2_2 - m^2_1}\,(m_\mu^2-m_e^2)
\,\bigg]\,\sin^2\theta\,\cos\theta\nn\\
&&\hspace{2cm}+\cos\theta\,m_1^2¡\,\Bigg\}\,,\\
\label{RGE_Ueff_e1}
\frac{d \left( {U^{\rm{eff}}_{CC}}\right)_{e2} }{d\ln\mu}  &=& 
\frac{g^2}{32 \pi^2 M_W^2}\,\Bigg\{
\,\frac{3}{2}\,
\bigg[\frac{ m^2_{\mu}+ m^2_e }{m^2_{\mu} - m^2_e}\,
(m_2^2-m_1^2) +
\frac{ (m_1+ m_2)^2 }{m^2_2 - m^2_1}\,(m_\mu^2-m_e^2)
\,\bigg]\,\sin\theta\,\cos^2\theta\nn\\
&&\hspace{2cm}-\sin\theta\,m_2^2
\,\Bigg\}\,.
\,
\label{RGE_Ueff_e2}
\eea

\subsection{RGE of $U^{\rm{eff}}_{NC}$}

From the definition of $U^{\rm{eff}}_{NC}$ in \eq{U_eff_NC}, we obtain its RGE 
at ${\cal O}(1/M^{2})$ in terms of the running of the unitary matrix $V$ in Eq.~(\ref{RGEV}) and the running of $\lambda_{NC}^\nu$: 

\be
\label{RGE_UeffNC}
\frac{d U^{\rm eff}_{NC}}{d\ln\mu}  =  -U^{\rm eff}_{NC}\,T_\nu\,
-\frac{1}{2}\,\frac{d {\lambda_{NC}^\nu}} {d\ln\mu} V^\dag\, .
\ee
where $T_\nu$ is given by \eq{Tnu} and
\be
-\frac{1}{2}\,\frac{d \lambda^{\nu}_{NC}} {d\ln\mu} V^\dag= 
-\frac{v^2}{4}\,\frac{d c_{\phi\ell}^{(-)}}{d\ln\mu}\, V^\dag\,= 
-\frac{1}{16\pi^2}\,\frac{9}{4}\, \frac{g^2}{2M_W^2}\,V^\dagger \,(m_\nu)^2_{\rm diag,real}\, 
\ee

Thus, \eq{RGE_UeffNC} becomes
\bea
\frac{d \left(U^{\rm eff}_{NC}\right)_{\alpha i} }{d\ln\mu}  &=&
\frac{g^2}{32 \pi^2 M_W^2}\,\frac{3}{2}\,\Bigg\{ 
\sum\limits_{j\neq i}^{n} 
\frac{(U_{CC}^{\rm{eff}})_{\alpha j}}{m^2_j - m^2_i}\,
\bigg[\,(m^2_j + m^2_i)\, 
{V_{j\beta }}\,
m^2_\beta\,V_{i\beta }^*\,
+\: 2m_j m_i\,
V^*_{j\beta}\, m^2_\beta\,
V_{i\beta }\,
\,\bigg]\,\nn\\
&&\hspace{2cm}-\frac{3\,V_{i \alpha }^*}{2}\,m_i^2
\Bigg\}\ .
\,
\eea

For the example of two light lepton generations considered previously, the RGE are
\bea
\frac{d \left( {U^{\rm{eff}}_{NC}}\right)_{e1} }{d\ln\mu}  &=& 
-\frac{g^2}{32 \pi^2 M_W^2}\,\,\frac{3}{2}\,\Bigg\{
\bigg[\frac{ m^2_{\mu}+ m^2_e }{m^2_{\mu} - m^2_e}\,
(m_2^2-m_1^2) +
\frac{ (m_2+ m_1)^2 }{m^2_2 - m^2_1}\,(m_\mu^2-m_e^2)
\,\bigg]\,\sin^2\theta\,\cos\theta\nn\\
&&\hspace{2.5cm}+\,\frac{3}{2}\,\cos\theta\,m_1^2¡\,\Bigg\}\,,\nn\\
\label{RGE_Ueff_e1_NC}
\frac{d \left( {U^{\rm{eff}}_{NC}}\right)_{e2} }{d\ln\mu}  &=& 
\frac{g^2}{32 \pi^2 M_W^2}\,\,\frac{3}{2}\,\Bigg\{
\bigg[\frac{ m^2_{\mu}+ m^2_e }{m^2_{\mu} - m^2_e}\,
(m_2^2-m_1^2) +
\frac{ (m_1+ m_2)^2 }{m^2_2 - m^2_1}\,(m_\mu^2-m_e^2)
\,\bigg]\,\sin\theta\,\cos^2\theta\nn\\
&&\hspace{2.5cm}-\,\frac{3}{2}\,\sin\theta\,m_2^2
\,\Bigg\}\,.
\,
\label{RGE_Ueff_e2_NC}
\eea

\section{RGE in the High-Energy Seesaw Theory}

Consider now the full seesaw Lagrangian in \eq{Lag}. The results in Ref.~\cite{PK} obtained with $n'$ right-handed neutrino  generations and $n$ lepton generations can be compared with the RGE obtained in the low-energy effective theory at \cal{O}$(1/M^{2})$.  We begin by reviewing the high energy theory. More details can be found in Appendix C.

After spontaneous breaking of the electroweak gauge symmetry, the portion of
the seesaw Lagrangian~(\ref{Lag}) which breaks the chiral symmetries of the lepton kinetic energy terms is given by
\bea
\label{Lag_SS}
{\cal L}_{\rm seesaw}^{\rm \chi SB}  &=& -\left[\frac{v}{\sqrt 2}\,\overline{{\nu}_{L}}\, Y_\nu\, N_{R} +
\frac{v}{\sqrt 2}\,\overline{e_L}\, Y_e\, e_R +\mrm {h.c.} \right] -
\frac{1}{2}\, \left( \overline{{N_{R}}^{c}} \,M\, N_{R} + \overline{N_R} M^* {N_R}^c \right) .
\eea

The left-handed neutrino fields can be arranged in a $(n + n^\prime)$ column vector
\be
n_L = 
\left(\begin{array}{c}
{\nu}_{L}\\
{N_{R}}^c
\end{array}\right)\, .
\ee
\eq{Lag_SS} can be written in terms of $n_L$ as 
\bea
{\cal L}_{\rm seesaw}^{\rm \chi S B} &=& - \left[\frac{1}{2}\, \overline{{n_{L}}^{c}} \,{\cal M} \,n_L +
\frac{v}{\sqrt 2}\,\overline{e_L}\, Y_e\, e_R \right] + \mrm{h.c.}\,,
\label{lm}
\eea
where the $(n+ n^\prime) \times (n + n^\prime)$ neutrino mass matrix ${\cal M}$ is defined 
in terms of the
$n \times n^\prime$ Yukawa coupling matrix $Y_\nu$ and the $n^\prime \times n^\prime$ Majorana 
mass matrix $M$ of the heavy neutrinos,
\bea
\label{cal_M2} 
{\cal M}\equiv \left(\begin{array}{cc}
0 & \frac{v}{\sqrt 2}\,Y_\nu^* \\
\frac{v}{\sqrt 2}\,Y_\nu^\dagger & M^* \end{array}\right) .
\eea
The neutrino mass matrix ${\cal M}$ is diagonalized by a $(n+ n^\prime) \times (n + n^\prime)$ unitary transformation ${\cal V}$,
\bea
{\cal V}^* {\cal M} {\cal V}^\dagger = \left( \cal M \right)_{\rm diag,\, real}\,,
\label{cal_diag}
\eea
where the diagonal mass matrix is described by its $(n+ n^\prime)$ mass 
eigenvalues.  Light eigenstates $\nu_i$  
get mass eigenvalues $m_i$, $i=1, \cdots, n$ and heavy eigenstates $N_{I}$ get masses $M_{I}$, ${ I}= n+1, \cdots, n+ n^\prime$.
The $n+ n^\prime$ Majorana mass eigenstate neutrinos $n=n^c$ are defined by
\bea
n \equiv 
\left(\begin{array}{c}
{\nu}\\
{N} \\
\end{array}\right)\,=
{\cal V} \ n_{L} + {\cal V}^* \ n_{L}^c . 
\eea

It is convenient to define submatrices of ${\cal M}_{\rm diag,\, real}$ and the diagonalizing matrix ${\cal V}$ \cite{Branco}:
\bea
\label{cal_D}
{\cal M}_{\rm diag,\, real}&=&\left(\begin{array}{cc}
m_{\rm diag,real} & 0 \\
0 & M_{\rm diag,\, real} \end{array}\right), \\ 
{\cal V} 
&\equiv& \left(\begin{array}{cc}
K & W \\
X & Z \end{array}\right) . 
\label{cal_V}
\eea

The neutrino weak eigenstates are
\be
\nu_{L \alpha} = P_L \left( \left(K^\dagger \right)_{\alpha i} \nu_i + \left( X^\dagger \right)_{\alpha I} N_{I} \right) ,
\ee
so that the weak currents are given by
\bea
&&J_\mu^{-\,CC} = \overline{e_{\alpha_L}}\, \gamma_{\mu}\, \big( E\,K^\dagger \big)_{\alpha i} \,
\nu_{i} +
\overline{e_{\alpha_L}}\, \gamma_{\mu} \, \big(E\,X^\dagger \big)_{\alpha I} \,{N_{I}} ,\\
\label{CC}
&&J_\mu^{NC}= \left({-{1 \over 2} -s^2\theta_W }\right)
\overline e_{L \alpha} \gamma_\mu e_{L \alpha}
+ \left(-s^2\theta_W \right) \overline e_{R \alpha} \gamma_\mu e_{R \alpha} \\
&&+ 
{1 \over 2} \,\big(\overline \nu_i \,\gamma_\mu
\,\big(K \, K^\dagger \big)_{ij} \, \nu_j + \overline N_{ I} \,\gamma_\mu\, \big(X \, X^\dagger \big)_{I J}\, N_J+\overline \nu_i \,\gamma_\mu
\,\big(K \, X^\dagger \big)_{iJ} \, N_J +  
\overline N_{ I} \,\gamma_\mu\, \big(X \, K^\dagger \big)_{Ij}
\, \nu_{j} \big)\,, \nn 
\label{NC}
\eea
where $E$ is the unitary matrix which diagonalizes the charged lepton mass matrix $m_e$. 

As our interest is in the renormalization of the weak currents which couple to the light neutrino eigenstates, we define the non-unitary matrices
\bea
\label{UCC}
&&U_{CC}\equiv E K^\dagger\\
\label{UNC}
&&U_{NC}\equiv K^\dagger
\eea
whose RGE will be compared below with the corresponding matrices obtained earlier in the effective theory at ${\cal O}(1/M^2)$. To this level of approximation, one can solve for the submatrices of ${\cal V}$ and ${\cal M}_{\rm diag, real}$ from the constraints that ${\cal V}$ be unitary and that it diagonalizes the mass matrix ${\cal M}$.  The complete analysis is given in Appendix \ref{appendixC}. The solution to ${\cal O}(1/M^2)$ for the mixing matrices is 
\bea
\label{K}
K &=& 
V \left( \mathbb I - \frac 1 2  \chi  \chi^\dagger \right) = V \left( \mathbb I - \frac \lambda 2 \right),  \\
\label{X}
X &=& -  V_H^* \chi^\dagger,
\eea
where the matrix $\chi$~\cite{bgj} is defined by
\be
\chi \equiv - {v \over \sqrt{2}} Y_\nu M^{-1} \,.
\ee

The matrix $V$ in \eq{K} is the same $n \times n$ unitary matrix which diagonalizes the light 
Majorana neutrino mass matrix at ${\cal O}(1/M^2)$, previously defined in 
\eq{diag_nu},
\bea
\label{mdr}
m_{\rm diag,\,real} &=& 
V^* \left(m_\nu \right) V^\dagger\, .
\eea
The matrix $V_H$ is the $n^\prime \times n^\prime$ unitary matrix which diagonalizes the complex, symmetric, heavy Majorana neutrino mass matrix at ${\cal O}(1/M^2)$,
\bea 
\label{Mdr}
M_{\rm diag,\,real} &=&  V_H \left( M^* +\frac {1} 2 \left(  \chi^T  \chi^* M^*  + M^*  \chi^\dagger \chi \right)\right) V_H^T \,.
\eea
Thus, to ${\cal O}(1/M^{2})$,
\bea
&&U_{CC}=EK^\dag = U^{\rm eff}_{CC}\,,\\
&&U_{NC}=K^\dag = U^{\rm eff}_{NC}\,,
\eea
with $U^{\rm eff}_{CC}$ and $U^{\rm eff}_{NC}$ being the mixing matrices of the effective theory, Eqs.~(\ref{U_eff_CC}) and~(\ref{U_eff_NC}).

The RGE of  $U_{CC}$ and $U_{NC}$ have been computed in Ref.~\cite{PK}  in the on-shell (OS) 
renormalization analysis of the CKM matrix \cite{Denner-Sack}. In order to compare the results in the full seesaw theory~\cite{PK} with the results in the effective theory, we expand the former 
to ${\cal O}(1/M^{2})$ using  Eqs.~(\ref{K}) and ~(\ref{X}). 
In the basis where $m_e$ and $M$ are real and diagonal, 
the RGE of $U_{CC}$ is given by
\bea\label{RGE_CC_comp}
\frac{d (U_{CC})_{ \alpha i}}{d\ln\mu}  &=& 
\frac{d \left(U^{\rm eff}_{CC}\right)_{\alpha i}}{d\ln\mu} 
+\theta(\mu-M)\,\frac{1}{16\pi^2}\,\frac{1}{2}\,\frac{g^2}{2M^2_W}\,V^*_{i \alpha}\,\,m^2_i \\
&+&\theta(M-\mu)\,\frac{1}{16\pi^2}\,\bigg\{ -\,\frac{1}{2}\,\lambda_{\alpha\beta}\,(Y_\nu\,Y_\nu^\dag)_{\beta\gamma}\, V^*_{i \gamma }\,+\frac{3}{2}\,\sum\limits^{n}_{\beta\neq \alpha}\,
\frac{ m^2_{\beta}+ m^2_\alpha }{m^2_{\beta} - m^2_\alpha}\,
\,(Y_\nu Y_\nu^\dag)_{\alpha \beta}
\,(U_{CC})_{ \beta i} \nn\\
&+& \frac{3}{2}\,\sum\limits_{j\neq i}^{n} 
\frac{(U_{CC})_{ \alpha j}}{m^2_j - m^2_i}\,
\bigg[\,(m^2_j + m^2_i)\, 
V_{j \beta }\,(Y_\nu 
Y_\nu^\dag)_{\beta\gamma}
\, V^*_{i \gamma}\,
+\, 2m_j m_i\, 
V^*_{j \beta }\,(Y_\nu Y_\nu^\dag)^*_{\beta\gamma}
\, V_{i \gamma}\,
\,\bigg]\,\,\bigg\}\,,\nn
\eea
and the RGE of $U_{NC}$ is
\bea\label{RGE_NC_comp}
\frac{d (U_{NC})_{ \alpha i}}{d\ln\mu}  &=& 
\frac{d \left(U^{\rm eff}_{NC}\right)_{ \alpha i}}{d\ln\mu}
+\theta(\mu-M)\,\frac{1}{16\pi^2}\,\frac{7}{4}\,\frac{g^2}{2M^2_W}\,
V^*_{i \alpha}\,\,m^2_i \nn\\
&+&\theta(M-\mu)\,\frac{1}{16\pi^2}\,\bigg\{
-\frac{1}{2}\,\lambda_{\alpha\beta}\,
(Y_\nu\,Y_\nu^\dag)_{\beta\gamma}\, V^*_{i \gamma}\,\\
&+& \frac{3}{2}\,\sum\limits_{j\neq i}^{n} 
\frac{(U_{NC})_{ \alpha j}}{m^2_j - m^2_i}\,
\bigg[\,(m^2_j + m^2_i)\, 
V_{j \beta }\,(Y_\nu 
Y_\nu^\dag)_{\beta\gamma}
\, V^*_{i \gamma}\,
+\, 2m_j m_i\, 
V^*_{j \beta }\,(Y_\nu Y_\nu^\dag)^*_{\beta\gamma}
\, V_{i \gamma}\,
\,\bigg]\,\,\nn
\,\bigg\}\,\,.
\eea
The $\theta$ functions in the equations have been 
introduced {\it by hand} to single out those terms which are different in both theories. The differences between the two theories 
stem from diagrams with heavy neutrino eigenstates running in the loop. Depending on the effects below 
the scale $M$ at which the heavy neutrino eigenstates decouple, we can divide these diagrams into 
two classes:

\begin{itemize}
\item{\emph{Diagrams which are UV divergent in the full theory and zero in the unbroken effective theory}.} For instance, the term 
\be
\frac{3}{2}\,\sum\limits^{n}_{\beta\neq \alpha}\,
\frac{ m^2_{\beta}+ m^2_\alpha }{m^2_{\beta} - m^2_\alpha}\,
\,(Y_\nu Y_\nu^\dag)_{\alpha \beta}
\,V^*_{i \beta }\, ,
\ee
present in Eq.~(\ref{RGE_CC_comp}), stems from the diagram in the full theory in Fig.~3.

\begin{picture}(400,150)(-20,0)
\Line(-20,140)(-20,-10)
\Line(200,140)(200,-10)
\Line(400,140)(400,-10)
\Line(-20,140)(400,140)
\Line(-20,110)(400,110)
\Line(-20,20)(400,20)
\Line(-20,-10)(400,-10)
\Line(-20,110)(200,110)
\put(60,120){\bf{Full Theory}}
\put(250,120){\bf{Effective Theory}}
\ArrowLine(20,40)(60,40)
\put(10,30){${e_{L \,\beta}}\,(p)$}
\Line(60,40)(120,40)
\put(80,30){${N_I}$}
\ArrowLine(120,40)(160,40)
\put(140,30){${e_{L \,\alpha}}\,(p)$}
\DashArrowArc(90,40)(30,0,180){5}
\put(80,75){ $\phi^+$}
\put(40,0){$\frac{i}{16\pi^2}\,\frac{1}{\epsilon}\,(Y_\nu Y_\nu^\dag)_{\alpha\beta}\,\frac{\not p}{2}\,  $}
\ArrowLine(230,40)(290,40)
\put(250,30){${e_{L \,\beta}}$}
\ArrowLine(290,40)(350,40)
\put(310,30){${e_{L \,\alpha}}$}
\DashArrowArc(290,70)(30,-90,270){5}
\put(280,80){ $\phi^+_l$}
\BCirc(290,40){3}
\put(250,0){$\frac{ig^2}{16\pi^2}\,\frac{1}{\epsilon}\,\lambda_{\alpha\beta}\,\frac{\not p}{2}\,  $}
\put(20,-20){ {\rm \scriptsize{\bf{Figure 3: Matching self-energy diagrams in the full and effective theories}}}}
\end{picture}
\vspace{1cm} 

The corresponding diagram in the effective theory, where the heavy neutrino propagator is  
replaced by a non-local interaction given by the $O^{d=6}$ operator, vanishes 
in dimensional regularization in the unbroken theory.

 In a mass-dependent regularization scheme, the diagram is not zero; instead it is proportional to the square 
of the cutoff which cancels the $1/M^2$ factor of the $d=6$ operator. This apparent 
contradiction between mass-dependent and mass-independent treatments is solved when all 
orders of perturbation theory are resummed in the mass-dependent scheme, recovering the null result 
obtained in dimensional regularization. After EWSB, this diagram does contribute positively to 
the RGE evolution, however, as the figure indicates.
 
All contributions arising from diagrams of this type are proportional to $Y_\nu Y_\nu^\dag$. Another 
example is provided by the term 
\be
-\,\frac{1}{2}\,\lambda_{\alpha\beta}\,(Y_\nu\,Y_\nu^\dag)_{\beta\gamma}\,
V^*_{i \gamma}
\ee
in Eqs.~(\ref{RGE_CC_comp}) and (\ref{RGE_NC_comp}),
 which also vanishes in the unbroken effective theory, although it
contributes after EWSB at ${\cal O}(1/M^{3})$, see Fig.~4.

\begin{picture}(300,150)(-20,0)
\Line(-20,140)(-20,-10)
\Line(200,140)(200,-10)
\Line(400,140)(400,-10)
\Line(-20,140)(400,140)
\Line(-20,110)(400,110)
\Line(-20,20)(400,20)
\Line(-20,-10)(400,-10)
\Line(-20,110)(200,110)
\put(60,120){\bf{Full Theory}}
\put(250,120){\bf{Effective Theory}}
\ArrowLine(-10,70)(30,70)
\put(-10,60){$\nu_j\,(p)$}
\Line(30,70)(160,70)
\put(60,60){${N_J}$}
\DashLine(120,40)(120,70){5}
\put(116,38){$\times$}
\put(100,60){$\nu_k$}
\DashLine(160,40)(160,70){5}
\put(156,38){$\times$}
\put(135,60){${N_I}$}
\ArrowLine(160,70)(190,70)
\put(165,60){$\nu_i(p)$}
\DashArrowArc(60,70)(30,0,180){5}
\put(60,85){ $\phi^+$}
\put(20,0){$\frac{i}{16\pi^2}\,\frac{1}{\epsilon}\,(V\,Y_\nu Y_\nu^\dag\,V^\dag)_{ij}\,\frac{\not p}{2}\, $}
\ArrowLine(220,70)(280,70)
\put(220,60){$\nu_j\,(p)$}
\Line(280,70)(340,70)
\put(300,60){$\nu_k$}
\ArrowLine(340,70)(380,70)
\put(360,60){$\nu_i\,(p)$}
\DashArrowArc(280,86)(15,-90,270){5}
\put(270,85){ $\phi^+_l$}
\DashLine(320,40)(340,70){5}
\put(316,38){$\times$}
\DashLine(360,40)(340,70){5}
\put(356,38){$\times$}
\BCirc(280,70){3}
\BCirc(340,70){3}
\put(280,0){${\cal O}(M^{-3})$}
\put(20,-20){ {\rm \scriptsize{\bf{Figure 4: Matching self-energy diagrams in the full and effective theories}}}}
\end{picture}
\vspace{1cm} 

\item{\emph{Diagrams which are UV finite in the complete theory and diverge in the effective theory.}} 
In addition to the terms in $Y_\nu Y_\nu^\dag$ explained above,  the terms
\be
\frac{1}{16\pi^2}\,\frac{g^2}{2M_W^2}\,\,V_{\alpha i}\,m_i^2
\ee
have a different numerical weight in Eqs.~(\ref{RGE_CC_comp}) and (\ref{RGE_NC_comp}). 
The origin of 
this difference is the same as for the case above: the replacement of the heavy neutrino propagators 
by the non-local interactions below the decoupling scale. Here, the heavy neutrino propagator 
is inside a non-UV divergent loop and its removal increases the degree of divergence, yielding 
UV divergent terms which contribute to the RGE. An example is given by the diagram in Fig.~5 showing a contribution which is UV safe in the full theory, but which is divergent in the effective theory. The same occurs in Fig.~6 where only one of the four diagrams which contributes to the neutrino
self-energy diverges in the 
full theory: in the effective theory a factor $4$ appears as the four 
diagrams are all UV divergent after decoupling. By studying the possible combinations of external legs, 
one realizes that the first three diagrams contribute to the effective operator $O^{(-)}_{\phi\ell}$, 
which also contains couplings $\overline{\nu}\,e\,\phi^+\,\phi^{0*}$,
see Fig.~6. The last diagram in Fig.~7, which is UV-safe in the full theory, gives rise 
to a divergent coefficient for the effective operator $O^{(1)}_{\phi\ell}$. 

\begin{picture}(300,150)(-20,-20)
\label{C}
\Line(-20,120)(-20,-40)
\Line(175,120)(175,-40)
\Line(400,120)(400,-40)
\Line(-20,120)(400,120)
\Line(-20,90)(400,90)
\Line(-20,-10)(400,-10)
\Line(-20,-40)(400,-40)
\Line(-20,90)(175,90)
\put(60,100){\bf{Full Theory}}
\put(230,100){\bf{Effective Theory}}
\ArrowLine(0,40)(40,40)
\put(-10,50){${e_{L \,\beta}}\,(p)$}
\Line(40,40)(120,40)
\put(40,30){${N_J}$}
\DashLine(65,0)(65,40){5}
\put(57,-3){ $\times$}
\put(75,30){${\nu_k}$}
\DashLine(95,0)(95,40){5}
\put(87,-3){ $\times$}
\put(100,30){${N_J}$}
\put(130,50){${e_{\alpha}}\,(p)$}
\put(75,65){ $\phi^+$}
\DashArrowArc(80,40)(40,0,180){5}
\ArrowLine(120,40)(160,40)
\put(60,-30){${\rm UV\,safe} $}
\ArrowLine(200,40)(240,40)
\put(190,50){${e_{\beta}}\,(p)$}
\Line(240,40)(320,40)
\DashLine(240,0)(240,40){5}
\put(234,-3){ $\times$}
\put(280,30){${\nu_k}$}
\DashLine(320,0)(320,40){5}
\put(314,-3){ $\times$}
\ArrowLine(320,40)(360,40)
\put(340,50){${e_{L \,\alpha}}\,(p)$}
\put(275,65){ $\phi^+$}
\DashArrowArc(280,40)(40,0,180){5}
\BBoxc(240,40)(6,6)
\BBoxc(320,40)(6,6)
\put(240,-30){$ \frac{i}{16\pi^2}\,\frac{1}{\epsilon}\,(c_5^*c_5)_{\alpha\beta}\,\frac{\not p}{2} $}
\put(20,-50){ {\rm \scriptsize{\bf{Figure 5: Matching self-energy diagrams in the full and effective theories.}}}}
\vspace{1cm}
\end{picture}

\begin{picture}(300,500)(0,40)
\Line(-20,540)(-20,-10)
\Line(200,540)(200,-10)
\Line(450,540)(450,-10)
\Line(-20,540)(450,540)
\Line(-20,510)(450,510)
\Line(-20,-10)(450,-10)
\Line(-20,410)(200,410)
\Line(-20,380)(200,380)
\Line(-20,280)(200,280)
\Line(-20,250)(200,250)
\Line(-20,150)(200,150)
\Line(-20,120)(200,120)
\Line(-20,20)(450,20)
\put(60,520){\bf{Full Theory}}
\put(300,520){\bf{Effective Theory}}
\ArrowLine(0,460)(40,460)
\put(-10,450){${\nu_j}\,(p)$}
\Line(40,460)(120,460)
\put(40,450){${N_J}$}
\DashLine(30,420)(30,460){5}
\put(22,417){ $\times$}
\put(75,485){ $\phi^{0}$}
\DashArrowArc(80,460)(20,0,180){5}
\put(75,450){${\nu_k}$}
\DashLine(130,420)(130,460){5}
\put(122,417){ $\times$}
\put(110,450){${N_J}$}
\put(150,450){$\nu_i\,(p)$}
\ArrowLine(120,460)(160,460)
\put(-15,395){${\frac{i}{16\pi^2}\,\frac{1}{\epsilon}\,\left[V\,\left(Y_\nu\frac{1}{M}Y_\nu^T\right)\,
\left(Y_\nu^*\frac{1}{M^*}Y_\nu^\dag\right)\,V^\dag\right]_{ij} \,\frac{\not p}{2}\,}$}
\ArrowLine(0,330)(40,330)
\put(0,320){${\nu_j}\,(p)$}
\Line(40,330)(120,330)
\put(40,320){${N_J}$}
\DashLine(35,290)(35,330){5}
\put(28,287){ $\times$}
\put(70,310){${\nu_k}$}
\DashLine(90,290)(90,330){5}
\put(84,287){ $\times$}
\put(100,320){${N_J}$}
\put(140,320){${\nu_i}\,(p)$}
\put(80,365){ $\phi^{0}$}
\DashArrowArc(90,330)(30,0,180){5}
\ArrowLine(120,330)(160,330)
\put(60,260){${\rm UV\,safe} $}
\ArrowLine(0,200)(40,200)
\put(0,190){${\nu_j}\,(p)$}
\Line(40,200)(120,200)
\put(40,190){${N_J}$}
\DashLine(65,160)(65,200){5}
\put(58,157){ $\times$}
\put(75,190){${\nu_k}$}
\DashLine(125,160)(125,200){5}
\put(119,157){ $\times$}
\put(105,190){${N_J}$}
\put(140,190){${\nu_i}\,(p)$}
\put(60,235){ $\phi^{0}$}
\DashArrowArc(70,200)(30,0,180){5}
\ArrowLine(120,200)(160,200)
\put(60,130){${\rm UV\,safe} $}
\ArrowLine(0,70)(40,70)
\put(0,60){${\nu_j}\,(p)$}
\Line(40,70)(120,70)
\put(40,60){${N_J}$}
\DashLine(65,30)(65,70){5}
\put(57,25){ $\times$}
\put(75,60){${\nu_k}$}
\DashLine(95,30)(95,70){5}
\put(87,25){ $\times$}
\put(100,60){${N_J}$}
\put(140,60){${\nu_i}\,(p)$}
\put(72,95){ $\phi^{0}$}
\DashArrowArc(80,70)(40,0,180){5}
\ArrowLine(120,70)(160,70)
\put(60,0){${\rm UV\,safe} $}
\ArrowLine(250,280)(290,280)
\put(240,270){${\nu_j}\,(p)$}
\Line(290,280)(370,280)
\DashLine(290,240)(290,280){5}
\put(284,237){ $\times$}
\put(325,270){${\nu_k}$}
\DashLine(370,240)(370,280){5}
\put(364,237){ $\times$}
\ArrowLine(370,280)(410,280)
\put(390,270){${\nu_i}\,(p)$}
\put(330,330){ $\phi^{0}$}
\DashArrowArc(330,280)(40,0,180){5}
\BBoxc(290,280)(6,6)
\BBoxc(370,280)(6,6)
\put(290,0){$ 4\,\frac{i}{16\pi^2}\,\frac{1}{\epsilon}\,(c_5^*c_5)_{\alpha\beta}\,\frac{\not p}{2}$}
\put(20,-20){ {\rm \scriptsize{\bf{Figure 6: Matching self-energy diagrams in the full and effective theories} }}}
\end{picture}

\begin{picture}(300,300)(0,-50)
\label{B}
\Line(-20,250)(-20,-40)
\Line(200,250)(200,-40)
\Line(400,250)(400,-40)
\Line(-20,250)(400,250)
\Line(-20,220)(400,220)
\Line(-20,-10)(400,-10)
\Line(-20,-40)(400,-40)
\Line(-20,90)(200,90)
\Line(-20,220)(200,220)
\Line(-20,120)(200,120)
\put(60,230){\bf{Full Theory}}
\put(260,230){\bf{Effective Theory}}
\ArrowLine(0,170)(40,170)
\put(-10,180){${\nu_j}\,(p)$}
\Line(40,170)(120,170)
\put(40,160){${N_J}$}
\DashLine(30,130)(30,170){5}
\put(22,127){ $\times$}
\put(75,195){ $\phi^{0}$}
\DashArrowArc(80,170)(20,0,180){5}
\put(75,160){${\nu_k}$}
\DashArrowLine(130,130)(130,170){5}
\put(135,130){ $\phi^{-}$}
\put(110,160){${N_J}$}
\put(130,180){$e_\alpha\,(p)$}
\ArrowLine(120,170)(160,170)
\put(-15,100){${\frac{-i}{16\pi^2}\,\frac{1}{\epsilon}\,\left[\left(Y_\nu\frac{1}{M}Y_\nu^T\right)\,
\left(Y_\nu^*\frac{1}{M^*}Y_\nu^\dag\right)\,V^\dag\right]_{\alpha j} \,\frac{\not p}{2}\, C_{\rm UV} }$}
\ArrowLine(0,40)(40,40)
\put(-10,50){${\nu_j}\,(p)$}
\Line(40,40)(120,40)
\put(40,30){${N_J}$}
\DashLine(65,0)(65,40){5}
\put(58,-3){ $\times$}
\put(75,30){${\nu_k}$}
\DashArrowLine(125,0)(125,40){5}
\put(130,0){$\phi^-$}
\put(105,30){${N_J}$}
\put(140,50){${e_{L \,\alpha}}\,(p)$}
\put(60,75){ $\phi^{0}$}
\DashArrowArc(70,40)(30,0,180){5}
\ArrowLine(120,40)(160,40)
\put(60,-30){${\rm UV\,safe} $}
\ArrowLine(210,90)(250,90)
\put(210,100){${\nu_j}\,(p)$}
\Line(290,90)(370,90)
\DashLine(250,50)(250,90){5}
\put(244,47){ $\times$}
\ArrowLine(250,90)(330,90)
\put(285,80){${\nu_k}$}
\DashArrowLine(330,50)(330,90){5}
\put(335,50){ $\phi^{-}$}
\ArrowLine(330,90)(370,90)
\put(350,100){${e_{L \,\alpha}}\,(p)$}
\put(280,140){ $\phi^{0}$}
\DashArrowArc(290,90)(40,0,180){5}
\BBoxc(250,90)(6,6)
\BBoxc(330,90)(6,6)
\put(250,-30){$ \frac{-i}{16\pi^2}\,\frac{1}{\epsilon}\,2\,(c_5^*c_5)_{\alpha\beta}\,\frac{\not p}{2}\,$}
\put(20,-50){ {\rm \scriptsize{\bf{Figure 7: Matching self-energy diagrams in the full and effective theories}}}}
\end{picture}
\vspace{1cm}
\end{itemize}

\section{Conclusions}

We have studied the renormalization of the leptonic mixing matrix in the case 
of light Majorana neutrinos.  The analysis has been performed in the low-energy effective theory
to ${\cal O}(1/M^{2})$, with $M$ being the generic high-energy  scale at which 
lepton number is broken. 
The leading flavor-changing RGE have been obtained including one-loop corrections which involve 
two insertions of the  $d=5$ neutrino mass operator as well as tree-level contributions of 
$d=6$ effective operators. 

In the existing derivations of the RGE at  ${\cal O}(1/M)$, the matrix which diagonalizes 
the $d=5$ operator coefficient in the unbroken theory also diagonalizes the 
neutrino mass matrix, and thus is the physical mixing matrix (when working 
in the basis in which the charged lepton Yukawa couplings are diagonal, as is customary).
In contrast, one-loop effects at ${\cal O}(1/M^{2})$ corresponding to two insertions of 
the $d=5$ operator coefficient result in flavor-changing corrections to 
the charged-lepton mass matrix. Its diagonalization also affects the evolution of the 
physical leptonic mixing matrix, 
just as in the quark case. This realization is crucial to the comparison of the effective 
theory results with the renormalization program of high-energy theories in which both heavy and light neutrino fields 
are active.

Furthermore, the tree-level and one-loop insertions of the $d=6$ operator characteristic of the  
seesaw theory yields non-unitary contributions to the mixing matrices of the effective theory, 
for both charged leptonic and neutrino neutral currents. 

We have finally compared our results in the effective theory with the RGE of the 
mixing matrices of the full seesaw theory developed to ${\cal O}(1/M^{2})$. 
The terms which differ in both expansions have been understood consistently 
in terms of the different 
degree of divergence of the diagrams in both theories.
\newpage

\section{Acknowledgments}
The work of A.B. and M.B.G. was partially supported by CICYT 100111003-54ID000-640 project.  A.B. acknowledges MECD for financial support through FPU fellowship AP2001-0521.  E.J. was supported in part by the Department of Energy under grant DE-FG03-97ER40546. 

\appendix
\section{Feynman rules for $d=5$ and $d=6$ operators}
\noindent

The $d=5$ effective Lagrangian for the seesaw model is given by
\bea
{\cal {\delta L}}^{d=5} &=& \frac{1}{2} (c_5)_{\alpha \beta} \left( \overline{{\ell_L}^c_\alpha}
 \tilde \phi^* \right) \left( \tilde \phi^\dagger \ell_{L\beta} \right) 
 + \mrm{h.c.}\nn\\
&=& \frac{1}{2} (c_5)_{\alpha \beta} \left( \overline{{\ell_L}^c_\alpha}
 \tilde \phi^* \right) \left( \tilde \phi^\dagger \ell_{L\beta} \right) 
 + \frac{1}{2} (c_5^*)_{\alpha \beta} \left( \overline{{\ell_L}_\alpha} \tilde \phi \right) \left( \tilde \phi^T {\ell_L}_\beta^c \right) \nn\\
&=&
\frac{1}{2}\, (c_{5})_{\alpha\beta}\,
\left({\overline{{\ell_L}^c_\alpha}}_{\,a}\,\epsilon^{am}\,  {\phi}_{m} \right)\,  
\left(\phi^T_n ({\epsilon^T})^{nb} {{\ell_L}_\beta}_b\right) + 
\frac{1}{2} (c_5^*)_{\alpha \beta} \left( \overline{{\ell_L}_\alpha}_a \epsilon^{am} \phi^*_m \right) \left(  \phi^\dagger_n (\epsilon^T)^{nb} {{\ell_L}_\beta^c}_b \right) \nn\\
&=&
\frac{1}{2}\, (c_{5})_{\alpha\beta}\,
\left({\overline{{\ell_L}^c_\alpha}}_{\,a}\,\epsilon^{am}\,  {\phi}_{m} \right)\,  
\left({{\ell_L}_\beta}_b \,\epsilon^{bn}\,{\phi}_{n} \right)+ \frac{1}{2} (c_5^*)_{\alpha \beta} \left( \overline{{\ell_L}_\alpha}_a \epsilon^{am} \phi^*_m \right) \left(  {{\ell_L}_\beta^c}_b  \epsilon^{bn}  \phi^*_n \right) \nn
\eea
where  $a,b,m,n$ are $SU(2)_L$ gauge indices, and $\epsilon=i\,\tau_2$ is the totally antisymmetric tensor in two dimensions. The Feynman rule for the $d=5$ vertex is:

\begin{picture}(0,100)(-100,0)
\ArrowLine(0,0)(40,40)
\ArrowLine(10,20)(20,30)
\put(-20,17){ ${\ell_{L \, \beta}}_b$}
\ArrowLine(80,0)(40,40)
\ArrowLine(60,30)(70,20)
\put(70,17){ ${\ell_{L\,\alpha}}_a$}
\DashArrowLine(0,80)(40,40){5}
\put(-10,60){ $\phi_n$}
\DashArrowLine(80,80)(40,40){5}
\put(70,60){ $\phi_m$}
\BBoxc(40,40)(6,6)
\put(100,40){ $i\,(c_{5})_{\alpha\beta}
\left(\epsilon^{am}\,\epsilon^{bn}+\epsilon^{an}\,\epsilon^{bm}\right)\, \left(\frac{1-\gamma_5}{2}\right)\,$}
\end{picture}
\vspace{1cm}

The extra arrows next to lepton lines indicate fermion flow, as defined in Ref.~\cite{Denner}

\begin{picture}(0,20)(-50,0)
\ArrowLine(30,10)(10,10)
\ArrowLine(0,0)(40,0)
\put(5,-15){ $\ell (k)$}
\put(50,-5){ $=$}
\ArrowLine(100,10)(80,10)
\ArrowLine(110,0)(70,0)
\put(75,-15){ $\ell (-k)$}
\put(120,-5){ $= -\frac{i}{\not k}$}
\end{picture}
\vspace{1cm}

The factor of $1/2$ in the Lagrangian term is cancelled by combinatoric
factor of $2$ for $\alpha \leftrightarrow \beta$ of the two lepton lines.  The two factors of $\epsilon \epsilon$ gives the
combinatoric factor of 2 for two $\phi$ lines.  The coefficient $c_5$ satisfies $c_5^T = c_5$.  The vertex is $\Delta L = -2$. 

The Hermitian conjugate $d=5$ vertex is:

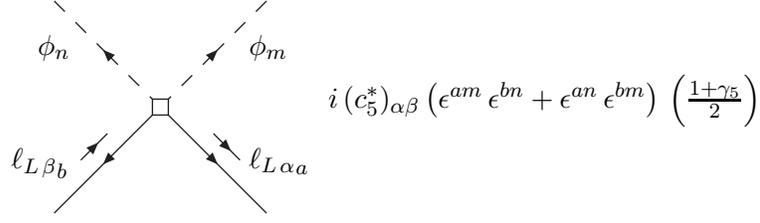
\begin{picture}(0,100)(-100,0)
\ArrowLine(40,40)(0,0)
\ArrowLine(10,20)(20,30)
\put(-20,17){ ${\ell_{L \, \beta}}_b$}
\ArrowLine(40,40)(80,0)
\ArrowLine(60,30)(70,20)
\put(70,17){ ${\ell_{L \,\alpha}}_a$}
\DashArrowLine(40,40)(0,80){5}
\put(-10,60){ $\phi_n$}
\DashArrowLine(40,40)(80,80){5}
\put(70,60){ $\phi_m$}
\BBoxc(40,40)(6,6)
\put(100,40){ $i\,(c_{5}^*)_{\alpha\beta}
\left(\epsilon^{am}\,\epsilon^{bn}+\epsilon^{an}\,\epsilon^{bm}\right)\, \left(\frac{1+\gamma_5}{2}\right)\,$}
\end{picture}
\vspace{1cm}

The vertex is $\Delta L = 2$.  Notice that chirality projection operator is now $P_R$.

The $d=6$ effective Lagrangian for the seesaw model is given by
\bea
{\cal {\delta L}}^{d=6} &=&   (c_{6})_{\alpha \beta}\,\left( \overline{\ell_{L \alpha}}\tilde \phi \right) \, i \dv \, 
\left(\tilde \phi^\dagger \ell_{L\beta} \right) \nn\\
&=& (c_{6})_{\alpha \beta}\ 
\left( \overline{{\ell_L}_\alpha}_{\,a}\,\epsilon^{am}\,  {\phi}^*_m\right) \,  i \dvr \left(   {{\ell_L}_\beta}_b\,\epsilon^{bn}\,{\phi}_{n} \right)\,.
\eea

The Feynman rule of the $d=6$ vertex is:

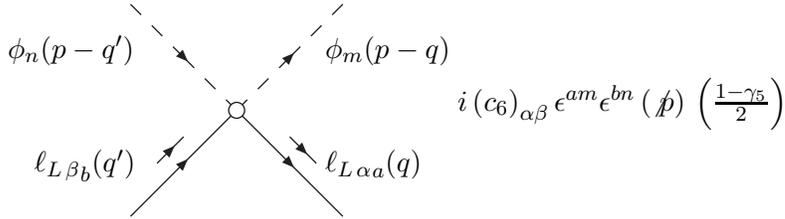
\begin{picture}(0,100)(-60,0)
\ArrowLine(0,0)(40,40)
\ArrowLine(10,20)(20,30)
\put(-40,17){ ${\ell_{L \,\beta}}_b(q^\prime)$}
\ArrowLine(40,40)(80,0)
\ArrowLine(60,30)(70,20)
\put(70,17){ ${\ell_{L \,\alpha}}_a(q)$}
\DashArrowLine(0,80)(40,40){5}
\put(-50,60){ $\phi_{n}(p-q^\prime)$}
\DashArrowLine(40,40)(80,80){5}
\put(70,60){ $\phi_{m} (p-q)$}
\BCirc(40,40){3}
\put(120,40){ ${i}\left(c_{6}\right)_{\alpha\beta}
\epsilon^{am}\epsilon^{bn}\left(\not p \right)\,\left(\frac{1-\gamma_5}{2}\right)$}
\end{picture}
\vspace{1cm}

The heavy right-handed
neutrino which was integrated out had momentum $p^\mu$.  The $c_6$ coefficient satisfies
$c_6^\dagger = c_6$.  The vertex is $\Delta L =0$.

\section{$|c_5|^2$ Diagrams}
\label{appendixB}

The BW $d=6$ operators induced by one-loop diagrams with two insertions of the $d=5$ operator are:
\bea
O^{(1)}_{\phi\ell}&=& i\left( {\overline\ell_L}_\alpha \,\gamma_\mu\,{\ell_L}_\beta \right ) \;
\left(\, {\phi}^\dagger\,D^\mu\,\phi\right)\,,\nn\\
O^{(3)}_{\phi\ell} &=& i \left( {\overline\ell_L}_\alpha \,\gamma_\mu\,\vec \tau\,{\ell_L}_\beta \right ) \;
\left(\, {\phi}^\dagger\,D^\mu\,\vec \tau\, \phi\right)\,,\nn\\
O^{(1)}_\phi &=& \left(\, {\phi}^\dagger\, \phi \right)\left(\, ({D_\mu  \phi}^\dagger)\,D^\mu \phi \right)\,,\nn\\
O^{(3)}_\phi &=&  \left(\, { \phi}^\dagger\, (D_\mu \phi) \right)\left(\,(D^\mu { \phi}^\dagger)\,\phi \right)\,,\nn\\
O_{e\phi}&=& (\overline\ell_L\,\phi\,e_R)\,\left(\, {\phi}^\dagger\phi \right)\,,\nn\\
O_{\ell\ell}^{(1)}&=&\frac{1}{2} \left(\overline{\ell_{L\alpha}}\gamma_\mu\ell_{L\beta}\right)\,
\left(\overline{\ell_{L\gamma}}\gamma^\mu\ell_{L\delta}\right)\nn\,.
\eea
and the Hermitian conjugate operators.

We will use the operator set:
\bea
O^{(-)}_{\phi \ell}&=& \frac{1}{2} \left( O^{(1)}_{\phi\ell} - O^{(3)}_{\phi\ell}\right) \equiv O_6,\nn\\
O^{(+)}_{\phi \ell}&=& \frac{1}{2} \left( O^{(1)}_{\phi\ell} + O^{(3)}_{\phi\ell}\right) ,\nn\\
O^{(1)}_\phi &=& \left(\, {\phi}^\dagger\, \phi \right)\left(\, ({D_\mu  \phi}^\dagger)\,D^\mu \phi \right)\,,\nn\\
O^{(3)}_\phi &=&  \left(\, { \phi}^\dagger\, (D_\mu \phi) \right)\left(\,(D^\mu { \phi}^\dagger)\,\phi \right)\,,\nn\\
O_{e\phi}&=& (\overline\ell_L\,\phi\,e_R)\,\left(\, {\phi}^\dagger\phi \right)\,,\nn\\
O_{\ell\ell}^{(1)}&=&\frac{1}{2} \left(\overline{\ell_{L\alpha}}\gamma_\mu\ell_{L\beta}\right)\,
\left(\overline{\ell_{L\gamma}}\gamma^\mu\ell_{L\delta}\right)\nn\,.
\eea
so that $O_6$ is an operator in the basis.

The effective Lagrangian is defined to be 
\be
{\cal L}_{\rm eff} \equiv \sum_i \, c_i \, O_i \ ,
\ee
which implies that the operator $O_i$ produces vertex with Feynman rule $i c_i S$, where $S$
is the symmetry factor of the operator.  (Equivalently, the operator $O_i/S$, defined with its proper
symmetry factor,  produces vertex with Feynman rule
$i c_i$.)

Coefficients $c_{\phi \ell}^{(-)} \equiv c_6$ and $c_{\phi \ell}^{(+)}$ are related to $c_{\phi \ell}^{(1)}$
and $c_{\phi \ell}^{(3)}$ by
\bea
c_{\phi \ell}^{(-)} &=& c^{(1)}_{\phi \ell} - c^{(3)}_{\phi \ell}\,, \nn\\
c_{\phi \ell}^{(+)} &=& c^{(1)}_{\phi \ell} + c^{(3)}_{\phi \ell}\nn
\eea
and
\bea
c_{\phi \ell}^{(1)} &=&  \frac{1}{2}\left( c^{(+)}_{\phi \ell} + c^{(-)}_{\phi \ell} \right)\,,\nn\\
c_{\phi \ell}^{(3)} &=& \frac{1}{2}\left( c^{(+)}_{\phi \ell} - c^{(-)}_{\phi \ell}  \right)\,. \nn
\eea

The UV parts of the one-loop diagrams with two insertions of the $d=5$ operator are calculated below.

\subsection{}
\label{1a}
\begin{picture}(0,100)(-50,0)
\label{d1}
\ArrowLine(0,0)(40,40)
\ArrowLine(10,20)(20,30)
\put(-50,10){${\ell_{L \,\beta}}_b\,(q^\prime)$}
\ArrowLine(80,40)(120,0)
\ArrowLine(100,30)(110,20)
\put(125,10){${\ell_{L \,\alpha}}_a\,(q)$}
\DashArrowLine(0,80)(40,40){5}
\put(-50,70){ $\phi_{n}\,(p-q^\prime)$}
\DashArrowLine(80,40)(120,80){5}
\put(125,70){ $\phi_{m}\,(p-q)$}
\BBoxc(40,40)(6,6)
\BBoxc(80,40)(6,6)
\DashArrowArc(60,40)(17,0,180){5}
\put(50,10){ ${\ell_{L\,\gamma}}_c$}
\ArrowLine(77,40)(43,40)
\ArrowLine(53,30)(67,30)
\put(50,67){ $\phi_l$}
\put(175,50){$=-\frac{i}{16\pi^2}\, \frac{1}{\epsilon} \,\frac{1}{2}\,  ({c_{5}}^*c_{5})_{\alpha\beta}
\left(\not p \right)\, \left(\frac{1-\gamma_5}{2} \right) $}
\put(185,20){$\times\,(4\,\epsilon^{am}\epsilon^{bn}+ \delta^{ab} \delta^{mn})$}
\end{picture}
\vspace{1cm}

The first term in parenthesis yields a contribution to the coefficient of $O_{6}$ 
\be
\Delta c_6= -\frac{1}{16\pi^2}\, \frac{1}{\epsilon}  \ 2\,\left({c_{5}}^*c_{5}\right)\,.
\ee

The second term in parenthesis gives rise to the following combination of $d=6$ operators:
\bea
\label{combi}
-\frac{i}{2}\left[i\left( {\overline\ell_L}_\alpha \,\gamma_\mu\,{\ell_L}_\beta \right ) 
\left(\, {\phi}^\dagger\,D^\mu\phi 
-(D^\mu{\phi}^\dagger)\,\phi \right)
+i\,\left( {\overline\ell_L}_\alpha \,\Dr\,{\ell_L}_\beta -{\overline\ell_L}_\alpha \,\Dl\,{\ell_L}_\beta \right ) \;
\left(\, \phi^\dagger \phi \right)\right]\,.
\eea

The first term is the operator $O^{(1)}_{\phi\ell}$
with a coefficient given by
\bea
\Delta c^{(1)}_{\phi\ell}&=& - \frac{1}{16\pi^2}\, \frac{1}{\epsilon} \ \frac{1}{4}\,\left({c_{5}}^*c_{5}\right)  
\eea
and it gives contributions to
\bea
\Delta c^{(+)}_{\phi\ell}&=& \Delta c^{(1)}_{\phi\ell} =
- \frac{1}{16\pi^2}\, \frac{1}{\epsilon} \ \frac{1}{4}\,\left({c_{5}}^*c_{5}\right)\,,\\  
\Delta c_{6}&=&  \Delta c^{(1)}_{\phi\ell} = - \frac{1}{16\pi^2}\, \frac{1}{\epsilon} \ \frac{1}{4}\,\left({c_{5}}^*c_{5}\right) \,.
\eea

Using the equations of motion of the lepton doublet 
\bea
\label{eqmot}
 i\, \Dr\,\ell_L\,-\,Y_e\,\phi\,e_R\,+\,
\cdots = 0 \,,\\
- i\,\overline{\ell_L}\, \Dl\,-\,\overline{e_R}\,\phi^\dagger\,Y_e^\dagger+
\cdots = 0 \,,
\eea
where the ellipsis stands for terms suppressed in $1/{M}$,
the second term generates 
$O_{e\phi}$
with a coefficient
\bea
\label{c1}
\Delta c_{e\phi}&=& - \frac{1}{16\pi^2}\, \frac{1}{\epsilon} \,\frac{1}{4}\,\left({c_{5}}^*c_{5}\;Y_e\right)\,.
\eea

\subsection{}
\label{1b}

\begin{picture}(100,100)(-40,0)
\label{d3}
\DashArrowLine(0,0)(40,40){10}
\put(-25,17){$\phi_{m}\,(p)$}
\DashArrowLine(80,40)(120,0){10}
\put(110,17){$\phi_l\,(p^\prime=p+q-q^\prime)$}
\DashArrowLine(0,80)(40,40){10}
\put(-25,60){ $\phi_{ n}\,(q)$}
\DashArrowLine(80,40)(120,80){10}
\put(110,60){ $\phi_k\,(q^\prime)$}
\BBoxc(40,40)(6,6)
\BBoxc(80,40)(6,6)
\ArrowArc(60,40)(17,0,180)
\ArrowArc(60,40)(27,75,105)
\put(57,77){${\ell_{L\,\alpha}}_a$}
\ArrowLine(77,40)(43,40)
\ArrowLine(53,30)(67,30)
\put(50,10){ ${\ell_{L \,\beta}}_b$}
\put(150,50){$=\frac{i}{16\pi^2}\, \frac{1}{\epsilon} \,{\rm Tr}[{c_{5}} {c_{5}}^*]\,
\left(p +q \right)^2 \left(\delta^{nl}\delta^{mk} + \delta^{nk}
\delta^{ml}\right) $}
\end{picture}
\vspace{0.5cm}

There is an explicit symmetry factor $S = \frac{1}{2}$ associated to this diagram. The diagram contributes to the following combination of operators 
\bea
&&-\left[\left(\phi^\dagger \phi \right) \left( \phi^\dagger \, D^2 \phi \right)
+ \left(\phi^\dagger \, D^2 \phi\right) \left( \phi^\dagger \phi \right)
+2\left(\, {\phi}^\dagger\,D^\mu\phi \right)\,
\left(\, {\phi}^\dagger\,D^\mu\phi \right)\right]= \nn\\
&&\qquad2\left(\, {\phi}^\dagger\,D^\mu\phi \right)
\left(\,(D^\mu {\phi}^\dagger)\,\phi \right)
+2\left(\, {\phi}^\dagger\,\phi \right)
\left(\,(D^\mu {\phi}^\dagger)\,D^\mu\phi \right)\nn\\
&&\qquad+\left(\phi^\dagger \, D^2 \phi \right) \left( \phi^\dagger \phi \right)
- \left(\phi^\dagger \phi \right) \left( \phi^\dagger \, D^2 \phi \right) \ . \nn
\eea
The first term on the right hand side of this equation is $2 O_\phi^{(3)}$.  The second one is $2 O_\phi^{(1)}$. The third and fourth terms on the right hand side can be rewritten using the equation of motion for $\phi$,
\be
D^2\phi = \overline{d_R}\,Y_d^\dagger\,q_L+\overline{q_L}\,\epsilon\,Y_u\,u_R
+\overline{e_R}\,Y_e^\dagger\,\ell_L\,
\ee
and shown to cancel.

Operators $O_\phi^{(1)}$ and $O_\phi^{(3)}$ result in the Feynman rule 
$2\left(\delta^{nl}\delta^{mk} + \delta^{nk}\delta^{ml}\right)$ because of a symmetry factor $S=4$. Considering all symmetry factors, the following coefficients are induced by the above diagram
\bea
\Delta c^{(1)}_\phi&=& \frac{1}{16\pi^2}\, \frac{1}{\epsilon}  \,{\rm Tr}[{c_{5}} {c_{5}}^*]\,,\\
\Delta c^{(3)}_\phi&=& \frac{1}{16\pi^2}\, \frac{1}{\epsilon} \,{\rm Tr}[{c_{5}} {c_{5}}^*]\,.
\eea

\subsection{}
\label{1c}
\begin{picture}(100,100)(-20,0)
\ArrowLine(0,0)(40,40)
\ArrowLine(10,20)(20,30)
\put(-20,17){${\ell_{L\,\delta}}_d$}
\ArrowLine(80,40)(120,0)
\ArrowLine(20,50)(10,60)
\put(120,17){${\ell_{L \,\beta}}_b$}
\ArrowLine(0,80)(40,40)
\ArrowLine(20,50)(10,60)
\put(-20,60){ ${\ell_{L\,\gamma}}_c$}
\ArrowLine(80,40)(120,80)
\ArrowLine(100,30)(110,20)
\ArrowLine(110,60)(100,50)
\put(120,60){ ${\ell_{L\,\alpha}}_a$}
\DashArrowArc(60,40)(17,0,180){5}
\put(50,20){ $\phi_n$}
\DashArrowLine(77,40)(43,40){5}
\put(50,67){ $\phi_m$}
\BBoxc(40,40)(6,6)
\BBoxc(80,40)(6,6)
\put(150,50){$=-\frac{i}{16\pi^2}\, \frac{1}{\epsilon}\,(c_{5}^*)_{\alpha\beta}\;({c_5})_{\gamma_\delta}\,    
\left(\overline{\ell_{L\beta}} \gamma^\mu \ell_{L\delta} \right) \left( \overline{\ell_{L\alpha}} \gamma_\mu \ell_{L\gamma}\right)$}
\end{picture}
\vspace{1cm}

Again, there is a symmetry factor $S=\frac{1}{2}$ and a symmetry factor $S=4$ in the Feynman rule for the operator $\left(\overline{\ell_{L\beta}} \gamma^\mu \ell_{L\delta} \right) \left( \overline{\ell_{L\alpha}} \gamma_\mu \ell_{L\gamma}\right)$.  Thus, the above diagram generates the coefficient

\be
\left(\Delta c_{\ell\ell}^{(1)}\right)_{\alpha\beta\gamma\delta}= -\frac{1}{16\pi^2}\, \frac{1}{\epsilon} \frac{1}{2}\,
(c_{5}^*)_{\alpha\gamma}\;(c_5)_{\beta\delta}\,, 
\ee
since
\be
\left(O_{\ell\ell}^{(1)}\right)_{\alpha\beta\gamma\delta} \equiv \frac{1}{2} \left(\overline{\ell_{L\alpha}} \gamma^\mu \ell_{L\beta}\right) \left(\overline{\ell_{L\gamma}} \gamma^\mu \ell_{L\delta}\right)
\ee
is defined with $1/2$.

\subsection{}
\label{1d}

\begin{picture}(300,140)(-20,0)
\label{d2}
\ArrowLine(10,10)(40,40)
\put(-10,25){ ${e_{R\,\beta}}$}
\ArrowLine(80,80)(120,30)
\ArrowLine(100,70)(110,57.5)
\put(115,45){ ${\ell_{L \,\alpha}}_a$}
\DashArrowLine(0,115)(40,80){5}
\put(-10,95){ $\phi_{k}$}
\DashArrowLine(30,125)(40,80){5}
\put(30,130){ $\phi_{l}$}
\DashArrowLine(80,80)(120,120){5}
\put(120,110){ $\phi_{m}$}
\ArrowLine(40,40)(40,80)
\ArrowLine(30,55)(30,65)
\put(-5,60){ ${\ell_{L\,\gamma}}_c$}
\ArrowLine(80,80)(40,80)
\ArrowLine(55,90)(65,90)
\put(50,100){ ${\ell_{L \,\delta}}_d$}
\DashArrowLine(80,80)(40,40){5}
\put(60,50){ $\phi_p$}
\BBoxc(40,80)(6,6)
\BBoxc(80,80)(6,6)
\put(150,80){$=\frac{i}{16\pi^2}\, \frac{1}{\epsilon} \,\,( {c_{5}}^*c_{5}\;Y_e)_{\alpha\beta}
\,(\delta^{ml} \delta^{ak}+\delta^{mk} \delta^{al})$}
\end{picture}

The above diagram contains two identical external $\phi$ lines, so there is a symmetry factor $S=2$ in
deducing its coefficient from its Feynman rule.  Thus, the above diagram induces the operator $O_{e\phi}$ with a coefficient
\bea
\label{c11}
\Delta c_{e\phi}&=&\frac{1}{16\pi^2}\, \frac{1}{\epsilon}  \,\left({c_{5}}^*c_{5}\,Y_e\,\right) \,.
\eea

Notice the difference with the case of diagram B1. There, we obtained $ O_{e\phi}$ from applying the 
equations of motion to another operator which does not contain identical particles.  Thus, we didn't apply a symmetry factor correction in Eq.~(7).

\subsection{}
In summary, the one-loop diagrams with two insertions of $c_5$ give a divergent contribution
to a number of the operator coefficients.   Specifically,

\bea
\Delta c_6&=& \Delta c_{\phi \ell}^{(-)} = - \frac{1}{16\pi^2}\, \frac{1}{\epsilon} \; \left( 2 + \frac{1}{4} \right)\,{c_{5}}^*c_{5}\,,\\
\Delta c^{(+)}_{\phi\ell}&=& - \frac{1}{16\pi^2}\, \frac{1}{\epsilon} \; \frac{1}{4}\,{c_{5}}^*c_{5}\,,\\
\Delta c_{e\phi}&=& \frac{1}{16\pi^2}\, \frac{1}{\epsilon} \;\frac{3}{4}\,
\left({c_{5}}^*c_{5}\;Y_e\right)\, ,\\
\Delta c^{(1)}_\phi&=& \frac{1}{16\pi^2}\, \frac{1}{\epsilon}\; {\rm Tr}[{c_{5}}^*c_{5}]\,,\\
\Delta c^{(3)}_\phi&=& \frac{1}{16\pi^2}\, \frac{1}{\epsilon}\; {\rm Tr}[{c_{5}}^*c_{5}]\,,\\
\left(\Delta c_{\ell\ell}^{(1)}\right)_{\alpha\beta\gamma\delta}&=& -\frac{1}{16\pi^2}\, \frac{1}{\epsilon} \; \frac{1}{2}\, (c_{5}^*)_{\alpha\gamma}\;
(c_5)_{\beta \delta}\, .
\eea
or
\bea
\Delta c^{(1)}_{\phi\ell}&=&- \frac{1}{16\pi^2}\, \frac{1}{\epsilon}\; \left( 1 + \frac{1}{4} \right)\,{c_{5}}^*c_{5}\,\nn\\
\Delta c^{(3)}_{\phi \ell} &=& \frac{1}{16\pi^2}\, \frac{1}{\epsilon}\; {c_{5}}^*c_{5}\,\nn 
\eea

\section{ High Energy Seesaw Theory}
\label{appendixC}

Considering \eq{cal_V}, unitarity of ${\cal V}$ implies
\bea
K K^\dagger +  W W^\dagger &=& \mathbb I, \\
X X^\dagger + Z Z^\dagger &=& \mathbb I, \\
X K^\dagger + Z W^\dagger &=& 0,\\
K^\dagger K + X^\dagger X &=& \mathbb I, \\
W^\dagger W + Z^\dagger Z &=& \mathbb I, \\
W^\dagger K + Z^\dagger X &=& 0,
\eea
so that the submatrices $K$, $W$, $X$ and $Z$ are not unitary.

Eq.~(\ref{cal_diag}) implies that
\bea
m_{\rm diag, \, real} &=& {v \over \sqrt{2}} \left( K^* Y_\nu^* W^\dagger + W^* Y_\nu^\dagger K^\dagger \right) + W^* M^* W^\dagger , \\
\label{zerocon}
0 &=&  {v \over \sqrt{2}} \left(  K^* Y_\nu^* Z^\dagger + W^* Y_\nu^\dagger X^\dagger  \right) + W^* M^* Z^\dagger ,\\
\label{Mcon}
M_{\rm diag, \, real} &=& {v \over \sqrt{2}} \left( X^* Y_\nu^* Z^\dagger + Z^* Y_\nu^\dagger X^\dagger \right) + Z^* M^* Z^\dagger, 
\eea

One can solve for the submatrices of ${\cal V}$ and ${\cal M}_{\rm diag, real}$ from the constraints that ${\cal V}$ be unitary and that it diagonalizes the mass matrix ${\cal M}$.  The solution to ${\cal O}(1/M^2)$ is
\bea
K &=& 
V \left( \mathbb I - \frac 1 2  \chi  \chi^\dagger \right) = V \left( \mathbb I - \frac \lambda 2 \right),  \\
\label{W}
W &=& V  \chi , \\
X &=& -  V_H^* \chi^\dagger,\\
\label{Z}
Z &=& V_H^* \left(  \mathbb I - \frac 1 2 \chi^\dagger \chi \right) ,
\eea
as well as Eqs.~(\ref{mdr},\ref{Mdr}).

The matrix ${\cal V}$ to ${\cal O}(1/M^2)$ is given by
\be
{\cal V} \equiv \left(\begin{array}{cc}
V \left( \mathbb I - \frac 1 2 \chi \chi^\dagger \right)  & V \chi \\
- V_H^*\chi^\dagger &  V_H^*\left( \mathbb I - \frac 1 2 \chi^\dagger \chi \right)
\end{array}\right)\,.
\ee


\end{document}